\documentclass[floatfix,twocolumn,showpacs,preprintnumbers,amsmath,amssymb,pra,superscriptaddress,longbibliography]{revtex4-1}
\usepackage{color}
\usepackage[usenames,dvipsnames,svgnames,table]{xcolor}
\usepackage[colorlinks=true,linkcolor=blue,urlcolor=blue,citecolor=blue]{hyperref}
\usepackage{mathtools}
\usepackage{graphicx}
\usepackage{dcolumn}
\usepackage{array}
\usepackage{lipsum}
\usepackage{bm}
\usepackage{subfigure}
\usepackage{amssymb}
\usepackage{multirow}
\usepackage{tabularx}
\usepackage{amsmath}
\usepackage{braket}
\usepackage{csquotes}
\graphicspath{{plots/}}
 \usepackage{lipsum}
\usepackage{mathrsfs}
\usepackage{MnSymbol}
	
\newcommand{\beq}{\begin{equation}}
\newcommand{\eeq}{\end{equation}}
\newcommand{\bea}{\begin{eqnarray}}
\newcommand{\eea}{\end{eqnarray}}

\begin{document}
\title{$\eta$-ensemble path integral Monte Carlo approach to the free energy of the warm dense electron gas and the uniform electron liquid}
\author{Tobias Dornheim}
\email{t.dornheim@hzdr.de}
\affiliation{Center for Advanced Systems Understanding (CASUS), D-02826 G\"orlitz, Germany}
\affiliation{Helmholtz-Zentrum Dresden-Rossendorf (HZDR), D-01328 Dresden, Germany}
\author{Panagiotis Tolias}
\affiliation{Space and Plasma Physics, Royal Institute of Technology (KTH), Stockholm, SE-100 44, Sweden}
\author{Zhandos~A.~Moldabekov}
\affiliation{Center for Advanced Systems Understanding (CASUS), D-02826 G\"orlitz, Germany}
\affiliation{Helmholtz-Zentrum Dresden-Rossendorf (HZDR), D-01328 Dresden, Germany}
\author{Jan Vorberger}
\affiliation{Helmholtz-Zentrum Dresden-Rossendorf (HZDR), D-01328 Dresden, Germany}
\begin{abstract}
We explore the recently introduced $\eta$-ensemble approach to compute the free energy directly from \emph{ab initio} path integral Monte Carlo (PIMC) simulations [T.~Dornheim \emph{et al.}, arXiv:2407.01044] and apply it to the archetypal uniform electron gas model both in the warm dense matter and strongly coupled regimes. Specifically, we present an in-depth study of the relevant algorithmic details such as the choice of the free weighting parameter and the choice of the optimum number of intermediate $\eta$-steps to connect the real, non-ideal system ($\eta=1$) with the ideal limit ($\eta=0$). Moreover, we explore the inherent decomposition of the full free energy into its ideal bosonic, ideal-to-interacting, and bosonic-to-fermionic contributions for different parameter regimes. Finally, we compare our new free energy data with an existing free energy parametrization [Groth \emph{et al.}, Phys.~Rev.~Lett.~\textbf{119}, 135001 (2017)] obtained via adiabatic connection formula evaluations, and we find very good agreement in its range of applicability, i.e., for density parameters $r_s\leq20$; in addition, we present the first PIMC results for the free energy in the low density regime of $20 < r_s\leq 100$. We expect our results to be of interest both for the study of matter under extreme conditions, as well as to the more general field of PIMC simulations of interacting quantum many-body systems.
\end{abstract}
\maketitle

\section{Introduction}

Understanding the properties of interacting Fermi systems at finite temperatures constitutes one of the most active fields within statistical, chemical, and atomic physics, with direct application to a gamut of other disciplines. Pertinent examples include ultracold quantum liquids such as fermionic $^3$He~\cite{Ceperley_PRL_1992,morresi2024normalliquid3hestudied} that is known to exhibit a wealth of interesting physics such as the BCS-BEC transition at low temperatures~\cite{dobbs2000helium} and a non-monotonic dispersion in the dynamic structure factor~\cite{Dornheim_SciRep_2022,Godfrin2012} that phenomenologically resembles the better known roton feature in the bosonic $^4$He~\cite{Trigger,Ferre_PRB_2016,Boninsegni1996}. Another example is given by electrons in quantum dots~\cite{Reimann_RevModPhys_2002}, which are known to form Wigner molecules~\cite{PhysRevB.62.8108,PhysRevB.63.113313,PhysRevLett.82.3320} and even to crystallize~\cite{Filinov_PRL_2001} at low densities, and which exhibit a counter-intuitive \emph{negative superfluid fraction} for certain numbers of electrons~\cite{Dornheim_NJP_2022,Yan_Blume_PRL_2014}. A particularly important example is given by so-called \emph{warm dense matter} (WDM)~\cite{wdm_book}, which is an extreme state that is characterized by high densities, temperatures, and pressures. These conditions are often defined in terms of two dimensionless parameters, which are both of the order of unity in the WDM regime~\cite{Ott2018}: i) the density parameter $r_s=d/a_\textnormal{B}$, where $d$ is the Wigner-Seitz radius and $a_\textnormal{B}$ the Bohr radius, and ii) the degeneracy temperature $\Theta=k_\textnormal{B}T/E_\textnormal{F}$, where $E_\textnormal{F}$ is the Fermi energy of the electrons~\cite{quantum_theory}. In nature, these conditions are abundant in astrophysical objects such as giant planet interiors~\cite{Benuzzi_Mounaix_2014}, brown dwarfs~\cite{becker} and white dwarf atmospheres~\cite{SAUMON20221}. Moreover, these conditions are important for the discovery and synthesis of materials, including the formation of nanodiamonds at high pressure and moderate temperatures first observed by Kraus and co-workers~\cite{Kraus2016,Kraus2017}. An especially important and topical application is given by inertial fusion energy (IFE)~\cite{Betti2016,Hurricane_RevModPhys_2023,drake2018high}, where both the fuel capsule and its surrounding ablator material have to traverse the WDM regime in a controlled way to reach ignition~\cite{hu_ICF}. Understanding the properties of WDM is thus of high importance to further optimize the recently reported positive energy gains achieved at the National Ignition Facility (NIF) at the Lawrence Livermore National Laboratory in California, USA~\cite{AbuShawareb_PRL_2024}, and to make IFE a commercially viable source of clean and safe energy~\cite{Batani_roadmap}.

Unfortunately, the theoretical description of WDM has proven notoriously difficult~\cite{wdm_book,new_POP,Dornheim_review}. Indeed, due to the absence of any small parameters as the basis for an expansion, one has to explicitly take into account the full interplay of thermal excitations, quantum effects, Coulomb coupling, and often partial ionization~\cite{Dornheim_Science_2024}. In practice, the required holistic description calls for sophisticated \emph{ab initio} methods, with the combination of thermal density functional theory (DFT) simulations~\cite{Mermin_DFT_1965} of the electrons with a molecular dynamics (MD) propagation of the ions being the workhorse of WDM theory~\cite{bonitz2024principles}. Still, DFT cannot be the starting point as it requires external input in the form of the electronic exchange--correlation (XC) functional; the latter is usually based on highly accurate quantum Monte Carlo (QMC) simulations of the uniform electron gas (UEG)---the archetypal system of interacting electrons~\cite{loos,quantum_theory,review}, and one of the most important model systems in statistical physics, quantum chemistry and related fields of research. 

At ambient conditions, the electrons are in their respective ground state and the key property that guides DFT simulations is given by the XC-energy. The latter is readily accessible from QMC simulations, which allows for systematic benchmarks of the zoo of existing XC-functionals~\cite{Goerigk_PCCP_2017,Clay_PRB_2014,Clay_PRB_2016}. At finite temperatures, one requires the XC-contribution to the \emph{free energy} $F=E-T\sigma$, where $E$ and $\sigma$ are the internal energy and entropy, respectively. The latter is equivalent to the partition function $Z$, to which it is related by
\begin{eqnarray}\label{eq:F}
    F = - \frac{1}{\beta} \textnormal{log} Z
\end{eqnarray}
with $\beta=1/k_\textnormal{B}T$ being the inverse temperature in energy units; $F$ can thus not directly be computed from \emph{ab initio} path integral Monte Carlo (PIMC) simulations~\cite{cep} and related finite-temperature QMC approaches~\cite{Schoof_PRL_2015,Malone_PRL_2016,Joonho_JCP_2021}.
Therefore, new thermal XC-functionals, which have been introduced over the last years~\cite{Sjostrom_PRB_2014,ksdt,groth_prl,Karasiev_PRL_2018,Karasiev_PRB_2022,kozlowski2023generalized}, cannot be directly assessed w.r.t~the free energy, and benchmarks have been limited to indirect comparisons of other observables such as the density~\cite{moldabekov2021relevance,Moldabekov_JCTC_2024} or the pressure~\cite{bonitz2024principles}.

Very recently, Dornheim \emph{et al.}~\cite{dornheim2024directfreeenergycalculation} partially changed this unsatisfactory situation by introducing a new scheme based on the connection between an ideal, uniform Bose gas [$\eta$=0, cf.~Eq.~(\ref{eq:Hamiltonian_eta})] and the real system of interest ($\eta=1$), which allows one to obtain the free energy of arbitrary quantum many-body systems. For PIMC simulations of WDM and other quantum degenerate Fermi systems, the method essentially comes at negligible computational overhead in addition to the usual fermionic sign problem~\cite{dornheim_sign_problem,troyer}. Moreover, it works equally well for non-uniform systems such as the electronic problem in the external potential of a fixed ion configuration---the standard problem in DFT-MD simulations of WDM---which is very difficult for hypothetically available alternative ans\"atze such as the adiabatic connection formula~\cite{Harding_JCP_2024}.

In the present work, we report an in-depth analysis of the algorithmic details of this $\eta$-ensemble scheme. Specifically, we carry out extensive PIMC simulations of the UEG, covering both the WDM and the strongly coupled electron liquid~\cite{dornheim_electron_liquid,Tolias_JCP_2021,castello2021classical,Tolias_JCP_2023,Dornheim_Force_2022} regime. Overall, we find a strong dependence of the optimal algorithmic parameters (i.e., number of intermediate $\eta$-steps $N_\eta$ and weighting parameter $c_{\eta}$) on the system parameters such as density parameter $r_s$, temperature $\Theta$, and also system size $N$. On the one hand, the estimation of the free energy in the moderately coupled WDM regime is relatively straightforward in practice since the number of required $\eta$-steps to connect the UEG with the ideal reference system is small. Instead, the main challenge here is given by the fermion sign problem that makes it difficult to resolve the quantum statistics correction given by the average sign $S$. On the other hand, the sign problem is very benign in the strongly coupled electron liquid regime where quantum degeneracy effects are markedly less important. Here, the main effort is given by comparably much larger number of required $\eta$-steps to switch between different coupling strengths in our simulations.

We expect our results to be of interest for a broad range of research fields. First and foremost, we present the first PIMC results for the free energy of the strongly coupled electron liquid. In addition to being interesting to their own right, these data constitute valuable benchmarks for the development of new methodologies such as dielectric theories~\cite{tanaka_hnc,Tanaka_CPP_2017,Tolias_JCP_2021,castello2021classical,Tolias_JCP_2023,Tolias_PRB_2024}. In addition, our study serves as a template for the future investigation of other warm dense matter systems such as the electronic problem in the external potential of a fixed ion configuration---the standard problem in DFT-MD simulations---or for two-component systems such as hydrogen~\cite{Dornheim_MRE_2024,bonitz2024principles,Militzer_PRE_2001} or beryllium~\cite{Dornheim_Science_2024,Dornheim_JCP_2024,dornheim2024modelfreerayleighweightxray}. Finally, the present approach is very general and it can be easily applied to PIMC simulations of a gamut of other quantum many-body systems such as ultracold atoms~\cite{cep,Filinov_PRA_2012,Dornheim_SciRep_2022}
or electrons in quantum dots~\cite{PhysRevLett.82.3320,Filinov_PRL_2001,dornheim_sign_problem}.

The paper is organized as follows: in Sec.~\ref{sec:theory}, we provide the theoretical background, including a brief introduction to the UEG and its Hamiltonian (Sec.~\ref{sec:UEG}), the PIMC method (\ref{sec:PIMC}) and its fermion sign problem (Sec.~\ref{sec:FSP}), the general idea of the extended ensemble sampling (Sec.~\ref{sec:extended_ensemble}), and the final result for the sought-after total free energy of the UEG (Sec.~\ref{sec:final}). Our new simulation results are presented in Sec.~\ref{sec:results} and include the in-depth investigation of the algorithmic efficiency in general (Sec.~\ref{sec:efficiency}) and specifically for the strongly coupled electron liquid (Sec.~\ref{sec:el}), as well as investigations of the XC-free energy $F_\textnormal{xc}$ (Sec.~\ref{sec:XC}) and the total free energy $F$ (Sec.~\ref{sec:total}).
The paper is concluded by a summary and outlook in Sec.~\ref{sec:outlook}.

\section{Theory\label{sec:theory}}

We assume Hartee atomic units throughout unless indicated otherwise.

\subsection{Uniform electron gas\label{sec:UEG}}

The UEG Hamiltonian is given by 
\begin{eqnarray}\label{eq:Hamiltonian}
    \hat{H}_\textnormal{UEG} = -\frac{1}{2}\sum_{k=1}^N \nabla^2_k + \frac{1}{2}\sum_{l,k=1;l\neq k}^N \Phi_\textnormal{E}(\hat{\mathbf{r}}_l,\hat{\mathbf{r}}_k)\ ,
\end{eqnarray}
with $\Phi_\textnormal{E}(\mathbf{r}_1,\mathbf{r}_2)$ being the usual Ewald pair potential that takes into account interactions between electrons, the positive uniform background, and with the infinite periodic array of images. Here, we follow the conventions as introduced in Ref.~\cite{Fraser_PRB_1996}, where $\Phi_\textnormal{E}(\mathbf{r}_1,\mathbf{r}_2)$ already includes the Madelung constant.

As $r_s\to0$ is approached, the kinetic energy predominates over the Coulomb coupling and the UEG attains the limit of an ideal Fermi gas. Conversely, the relative importance of Coulomb coupling starts to increase with $r_s$, and $r_s\sim10-100$ corresponds to the strongly coupled electron liquid regime~\cite{dornheim_electron_liquid,Tolias_JCP_2021,Tolias_JCP_2023,Takada_PRB_2016,quantum_theory,dornheim_dynamic}, which exhibits interesting effects such as an effective electron--electron attraction~\cite{Takada_PRB_1993,Dornheim_Force_2022,Takada_PRB_2024} and a \emph{roton-type} feature in the dynamic structure factor~\cite{Dornheim_Nature_2022,dornheim_dynamic,Takada_PRB_2016,koskelo2023shortrange}.
For even lower densities, the UEG will eventually form a Wigner crystal~\cite{Wigner_PhysRev_1934}, although the precise phase boundary is still being debated even at zero temperature~\cite{Drummond_PRB_Wigner_2004,Holzmann_PRL_2020,Azadi_PRB_2022}.

\subsection{Path integral Monte Carlo\label{sec:PIMC}}

The PIMC method has originally been introduced for the simulation of ultracold $^4$He~\cite{Fosdick_PR_1966,Jordan_PR_1968} and constitutes one of the most successful tools for the simulation of interacting quantum many-body systems at finite temperatures. Since an extensive and accessible review has been presented by Ceperley~\cite{cep}, we here restrict ourselves to a brief summary of the relevant details.

The basic idea is to express the canonical partition function (i.e., volume $\Omega=L^3$, number density $n=N/\Omega$, and inverse temperature $\beta$ are being kept constant) in the coordinate representation,
\begin{widetext}
\begin{eqnarray}\label{eq:coordinate_representation}
    Z(\Omega,n,\beta) = \frac{1}{N_\uparrow!N_\downarrow!} \sum_{\sigma_\uparrow\in S_{N_\uparrow}}\sum_{\sigma_\downarrow\in S_{N_\downarrow}} \textnormal{sgn}(\sigma_\uparrow,\sigma_\downarrow) \int_{\Omega^N} \textnormal{d}\mathbf{R}\ \bra{\mathbf{R}} e^{-\beta \hat{H}} \ket{\hat{\pi}_{\sigma_\uparrow}\hat{\pi}_{\sigma_\downarrow}\mathbf{R}}
\end{eqnarray}
\end{widetext}
where $\mathbf{R}=(\mathbf{r}_1,\dots,\mathbf{r}_N)^T$ contains the coordinates of all $N$ particles. Note that we restrict ourselves to the spin-unpolarized case of $N_\uparrow=N_\downarrow=N/2$ throughout.
To take into account the anti-symmetry of the thermal density matrix under the exchange of particle coordinates of electrons with the same spin orientation, one has to sum over all possible permutations $\sigma_i$ of the respective permutation groups $S_{N_i}$ [with $i\in\{\uparrow,\downarrow\}$], generated by the corresponding exchange operators $\hat{\pi}_{\sigma_i}$, where the sign is flipped for every pair exchange; this is taken into account by the sign function in front of the integral.

In practice, the matrix elements of the thermal density operator $\hat{\rho}=e^{-\beta\hat{H}}$ cannot be evaluated in general as the kinetic and potential contributions to the total Hamiltonian $\hat{H}=\hat{K}+\hat{V}$ do not commute. 
This is usually solved by evoking the well-known semi-group property of $\hat{\rho}$ and the subsequent Trotterization, i.e., factorization of the density matrix, which becomes exact in the limit of a large number of high-temperature factors $P$. Formally, one can interpret $\hat{\rho}$ as the equilibrium time evolution operator evaluated for an imaginary-time propagation of $t=-i\beta$, and the aforementioned Trotterization as a discrete imaginary-time grid of slice length $\epsilon=\beta/P$. The basic idea of the PIMC method is then to stochastically sample all possible imaginary-time path configurations $\mathbf{X}$ according to their appropriate configuration weight $W(\mathbf{X})$.
Schematically, we can express the partition function as
\begin{eqnarray}\label{eq:Z_dX}
Z = \sumint\textnormal{d}\mathbf{X}\ W(\mathbf{X})\ ,
\end{eqnarray}
where the symbolic notation $\sumint\textnormal{d}\mathbf{X}$ implies both the integration over all sets of particle coordinates $\mathbf{X}=(\mathbf{R}_0,\dots,\mathbf{R}_{P-1})^T$ and the sum over all possible permutations of particle coordinates, and where the associated unnormalized configuration weight $W(\mathbf{X})$ is a function that can be readily evaluated.

Modern sampling methods are generally based on the celebrated Metropolis algorithm~\cite{metropolis}, and efficient transitions between different permutation topologies are ensured by the worm algorithm~\cite{boninsegni1,boninsegni2} and canonical implementations thereof~\cite{mezza,Dornheim_PRB_nk_2021}.

\subsection{Fermion sign problem\label{sec:FSP}}

For bosons and hypothetical distinguishable quantum particles (often denoted as \emph{boltzmannons}), all contributions to Eq.~(\ref{eq:Z_dX}) are positive, and quasi-exact (i.e., exact within the given Monte Carlo error bars) results can be obtained for up to $N\sim10^4$ particles. For fermions, this is not the case due to the sign function in Eq.~(\ref{eq:coordinate_representation}). Consequently, we cannot interpret $P(\mathbf{X})=W(\mathbf{X})/Z$ as a probability distribution, which prevents straightforward Metropolis sampling.

To overcome this obstacle, we sample according to the modified weight function $W'(\mathbf{X})=|W(\mathbf{X})|$, with the corresponding modified normalization
\begin{eqnarray}\label{eq:modified_partition_function}
    Z' = \sumint\textnormal{d}\mathbf{X}\ |W(\mathbf{X})| = Z_B\ .
\end{eqnarray}
In the case of the direct PIMC method that we employ in this work, $Z'$ corresponds to the physical partition function of a Bose system at the simulated conditions, $Z'(\Omega,n,\beta) = Z_\textnormal{B}(\Omega,n,\beta)$. In other words, one extracts information about the Fermi system of interest from a bosonic reference PIMC simulation,
and it is easy to see that the fermionic expectation value of an observable $\hat{O}$ is given by
\begin{eqnarray}\label{eq:ratio}
    \braket{\hat{O}}=\frac{\braket{\hat{O}\hat{S}}'}{\braket{\hat{S}}'}\ ,
\end{eqnarray}
where $\braket{\dots}'$ denotes the expectation value computed in the modified configuration space.
In this context, a key property is given by the demoninator of Eq.~(\ref{eq:ratio}), to which we will refer as the \emph{average sign}. It is defined as the expectation value of the sign operator
\begin{eqnarray}
    S(\mathbf{X}) = \frac{W(\mathbf{X})}{|W(\mathbf{X})|}
\end{eqnarray}
evaluated in the modified configuration space,
\begin{eqnarray}\label{eq:S}
    S  = \braket{\hat{S}}' = \frac{1}{Z'}\sumint \textnormal{d}\mathbf{X}\ S(\mathbf{X})\ |W(\mathbf{X})| = \frac{Z}{Z'}\ ,
\end{eqnarray}
and constitutes a convenient measure for the degree of cancellation between positive and negative contributions to Eq.~(\ref{eq:coordinate_representation}) due to the fermionic antisymmetry~\cite{dornheim_sign_problem,troyer}.
In practice, fermionic PIMC simulations are generally feasible for $S\gtrsim10^{-2}\dots10^{-3}$~\cite{dornheim_sign_problem,Dornheim_JPA_2021}.
The sign is easily expressed as the ratio of fermionic and bosonic partition functions.
In combination with Eq.~(\ref{eq:F}), we can thus write
\begin{eqnarray}\label{eq:F_and_S}
    F_\textnormal{F} = F_\textnormal{B} - \frac{1}{\beta}\textnormal{log}\left(
S
    \right)\, ;
\end{eqnarray}
in other words, the free energy of the fermionic system of interest is given by the free energy of the corresponding Bose system that is defined in Eq.~(\ref{eq:modified_partition_function}) and a quantum statistics correction given by the average sign $S$. The latter is a standard observable in PIMC simulations and is indispensable for the evaluation of any fermionic expectation value, cf.~Eq.~(\ref{eq:ratio}). The task at hand is thus the estimation of the free energy of the non-ideal bosonic many-body system, which is discussed in what follows.

\subsection{Extended ensemble sampling\label{sec:extended_ensemble}}

\begin{figure}\centering
\includegraphics[width=0.37\textwidth]{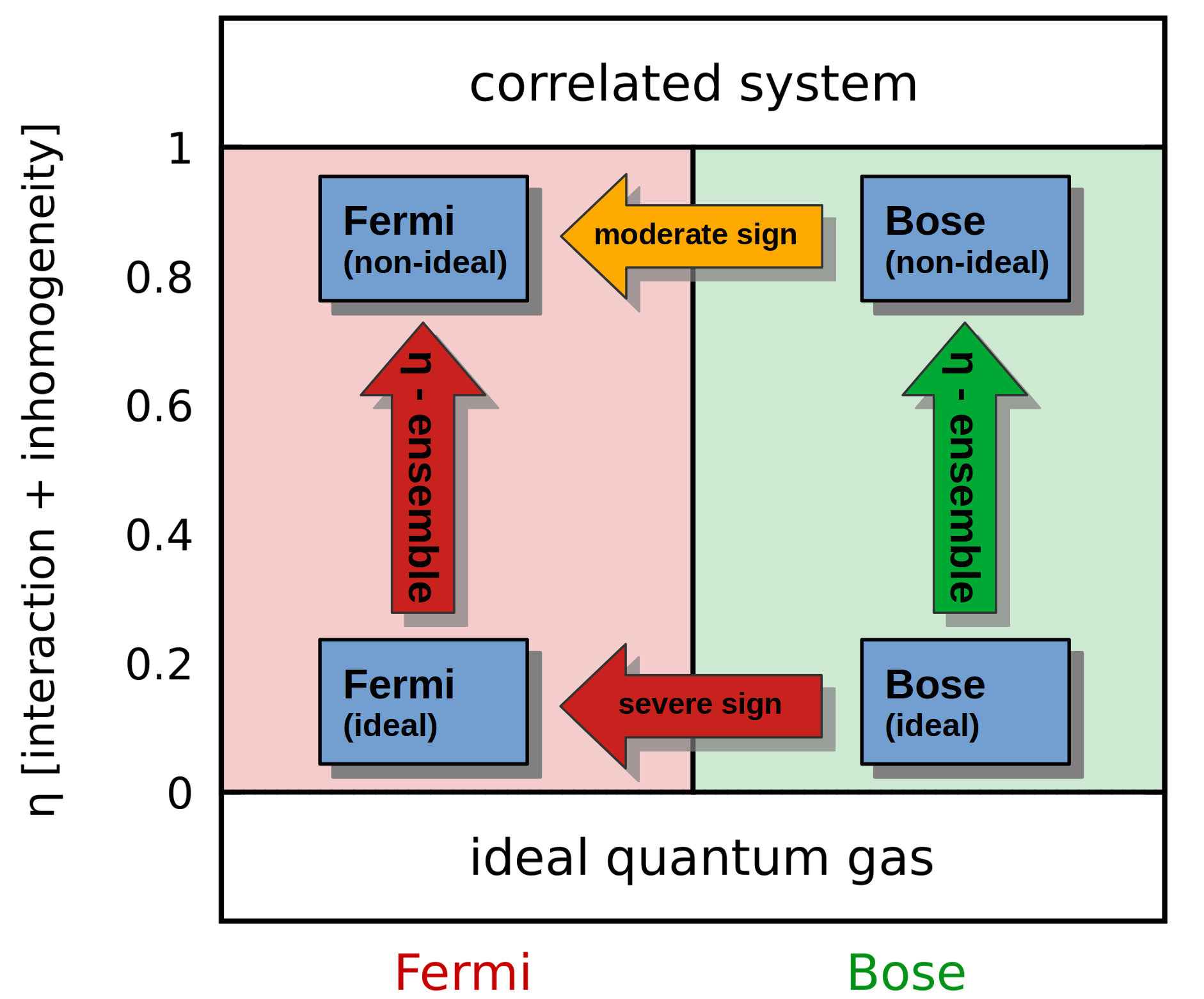}
\caption{\label{fig:scheme} Schematic illustration of the direct free energy estimation from PIMC simulations [cf.~Eq.~(\ref{eq:final})]. The direct route from the ideal Fermi gas to the interacting UEG of interest (left vertical arrow) is, in principle, possible, but subject to a severe sign problem due to the large prevalence of quantum degeneracy effects in the noninteracting system. To avoid this obstacle, we start at the ideal Bose gas (bottom left) and connect it with the interacting Bose reference system (right arrow), given the sign-problem free nature of bosonic systems. The final step is then given by the quantum statistical correction (top arrow) between the interacting systems, where the sign problem is substantially less severe. Taken from Ref.~\cite{dornheim2024directfreeenergycalculation}.
}
\end{figure} 

At this point, the only missing ingredient is the free energy of the non-ideal Bose system, for which also no general analytic expression exists. To overcome this obstacle, we define a modified Hamiltonian~\cite{dornheim2024directfreeenergycalculation}
\begin{eqnarray}\label{eq:Hamiltonian_eta}
    \hat{H}_\eta = \hat{K} + \eta \hat{V}\ ,
\end{eqnarray}
with $\eta\in[0,1]$ a free parameter. The true UEG Hamiltonian [Eq.~(\ref{eq:Hamiltonian})] is recovered for $\eta=1$, whereas the opposite limit of $\eta=0$ describes a noninteracting (i.e., ideal) Bose or Fermi gas. For the latter, the free energy is known both in the thermodynamic limit~\cite{quantum_theory} and also for the canonical ensemble with finite $N$ based on a well-known recursion relation of $Z_{\textnormal{B},0}(\Omega,n,\beta)$~\cite{Zhou_2018}, see Eq.~(\ref{eq:Z_ideal_Bose_Fermi}).

To connect the interacting and non-interacting Bose systems, we generalize the spirit behind Eq.~(\ref{eq:F_and_S}) and define an extended ensemble governed by the combined partition function
\begin{eqnarray}\label{eq:Z_extended}
    Z_{\eta_1,\eta_2} &=& c_{\eta_1}Z[\hat{H}_{\eta_1}]  +  c_{\eta_2} Z[\hat{H}_{\eta_2}] \\ 
    &=& \sum_{\eta\in\{\eta_1,\eta_2\}} c_{\eta} \sumint \textnormal{d}\mathbf{X}\ W(\mathbf{X},\eta)
\end{eqnarray}
and we set $c_{\eta_2}\equiv1$ throughout. The constant $c_{\eta_1}$ is a free, positive definite parameter that controls the relative weights of the two $\eta$-sectors; it can be chosen to improve the algorithmic efficiency as it will be explained below. Without loss of generality, we further assume that $\eta_1 > \eta_2$. The generalized configuration weight differs from the previously introduced $W(\mathbf{X})$ only in its interaction part,
\begin{eqnarray}\label{eq:}
    W(\mathbf{X},\eta) = W(\mathbf{X}) e^{-(\eta-1)\epsilon V(\mathbf{X})}\ ,
\end{eqnarray}
where $V(\mathbf{X})$ denotes the sum of Ewald interaction energies of all $P$ time slices.

To switch between the two $\eta$-sectors, which is required to sample the full configuration space defined by Eq.~(\ref{eq:Z_extended}), we have implemented an additional update into the \texttt{ISHTAR} code~\cite{ISHTAR}.
Specifically, we propose to switch from $\eta_1$ to $\eta_2$ (and vice versa) without any explicit modification of the path configuration $\mathbf{X}$, leading to the Metropolis~\cite{metropolis} acceptance probabilities of 
\begin{eqnarray}\label{eq:A1}
    A(\eta_1\to\eta_2) &=& \textnormal{min}\left\{
1, c_{\eta_1}^{-1} e^{-(\eta_2-\eta_1)\epsilon V(\mathbf{X})}
    \right\} \\
        A(\eta_2\to\eta_1) &=& \textnormal{min}\left\{
1, c_{\eta_1} e^{-(\eta_1-\eta_2)\epsilon V(\mathbf{X})}
    \right\}\ .\label{eq:A2}
\end{eqnarray}
The ratio of the two partition functions is then estimated based on the observable
\begin{eqnarray}\label{eq:delta}
    \hat{\delta}_{\eta_i}= 
\begin{cases}
    1,& \text{if } \mathbf{X} \in Z_{\eta_i}\\
    0,              & \text{otherwise}\ ,
\end{cases}
\end{eqnarray}
where $i\in\{\eta_1,\eta_2\}$;
it counts the frequency of the respective $\eta$-sectors in the full Markov chain.
This immediately leads to the partition function ratio
\begin{eqnarray}\label{eq:r}
    r(\eta_1,\eta_2) &=& \frac{\braket{\hat{\delta}_{\eta_1}}'}{\braket{\hat{\delta}_{\eta_2}}'} \\ 
    \Rightarrow \frac{Z[\hat{H}_{\eta_1}]}{Z[\hat{H}_{\eta_2}]} &=& \frac{r(\eta_1,\eta_2)}{c_{\eta_1}}\ , \label{eq:r2}
\end{eqnarray}
and, consequently, to the free energy expression
\begin{eqnarray}
    F_{\eta_1} - F_{\eta_2} = -\frac{1}{\beta}\textnormal{log}\left(
\frac{r(\eta_1,\eta_2)}{c_{\eta_1}}
    \right)\ .
\end{eqnarray}
The implementation of the extended ensemble is thus a convenient way to estimate the free energy difference between the two systems defined by $\hat{H}_{\eta_1}$ and $\hat{H}_{\eta_2}$~\cite{dornheim2024directfreeenergycalculation,Lyubartsev_JCP_1992}.

\begin{figure*}\centering
\includegraphics[width=0.32\textwidth]{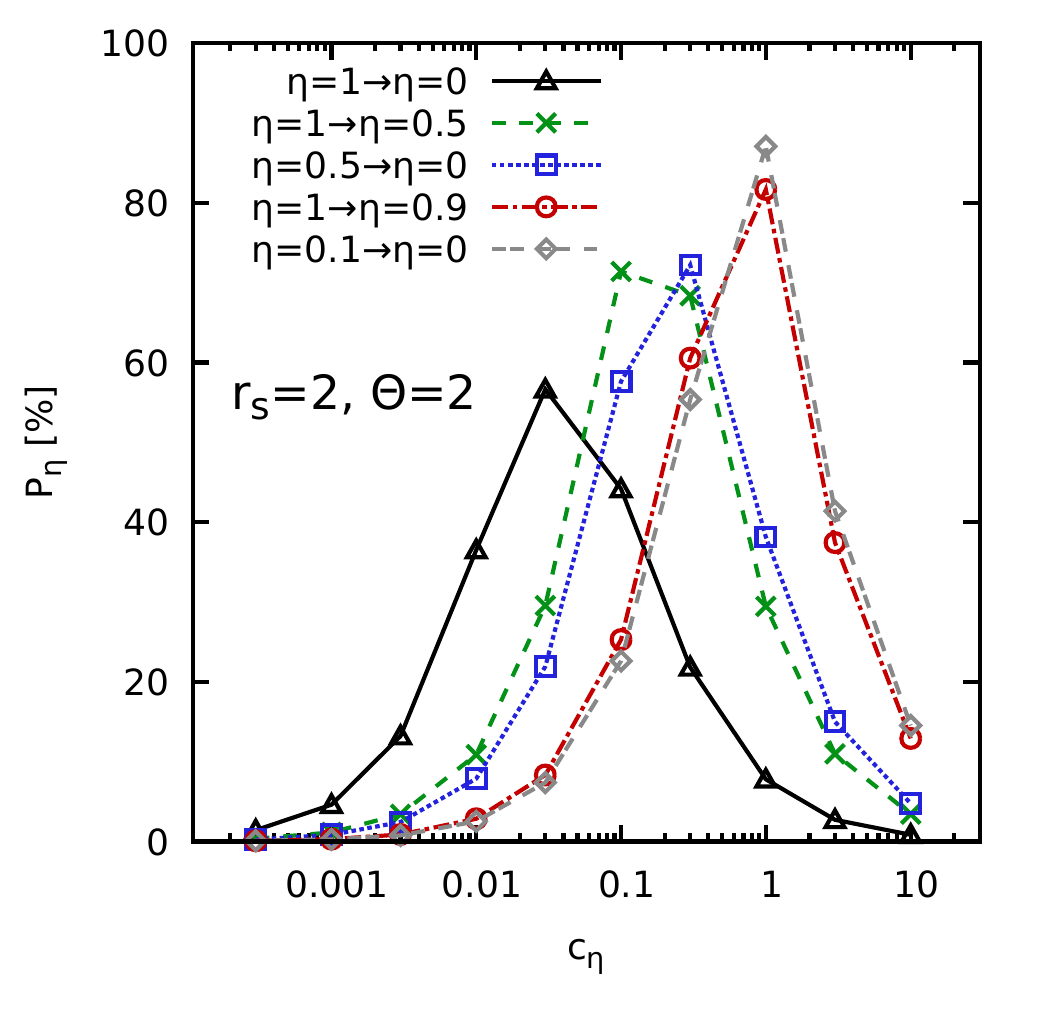}\includegraphics[width=0.32\textwidth]{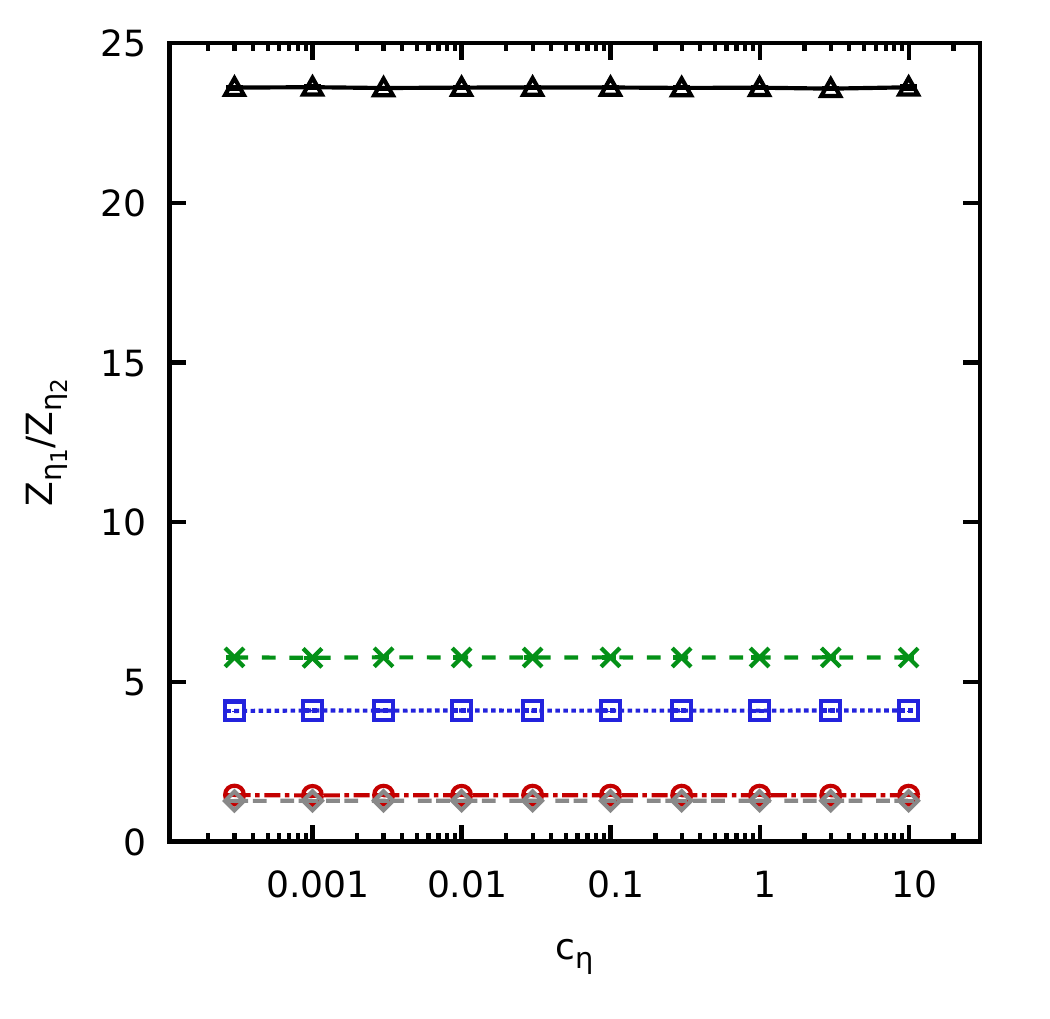}\includegraphics[width=0.32\textwidth]{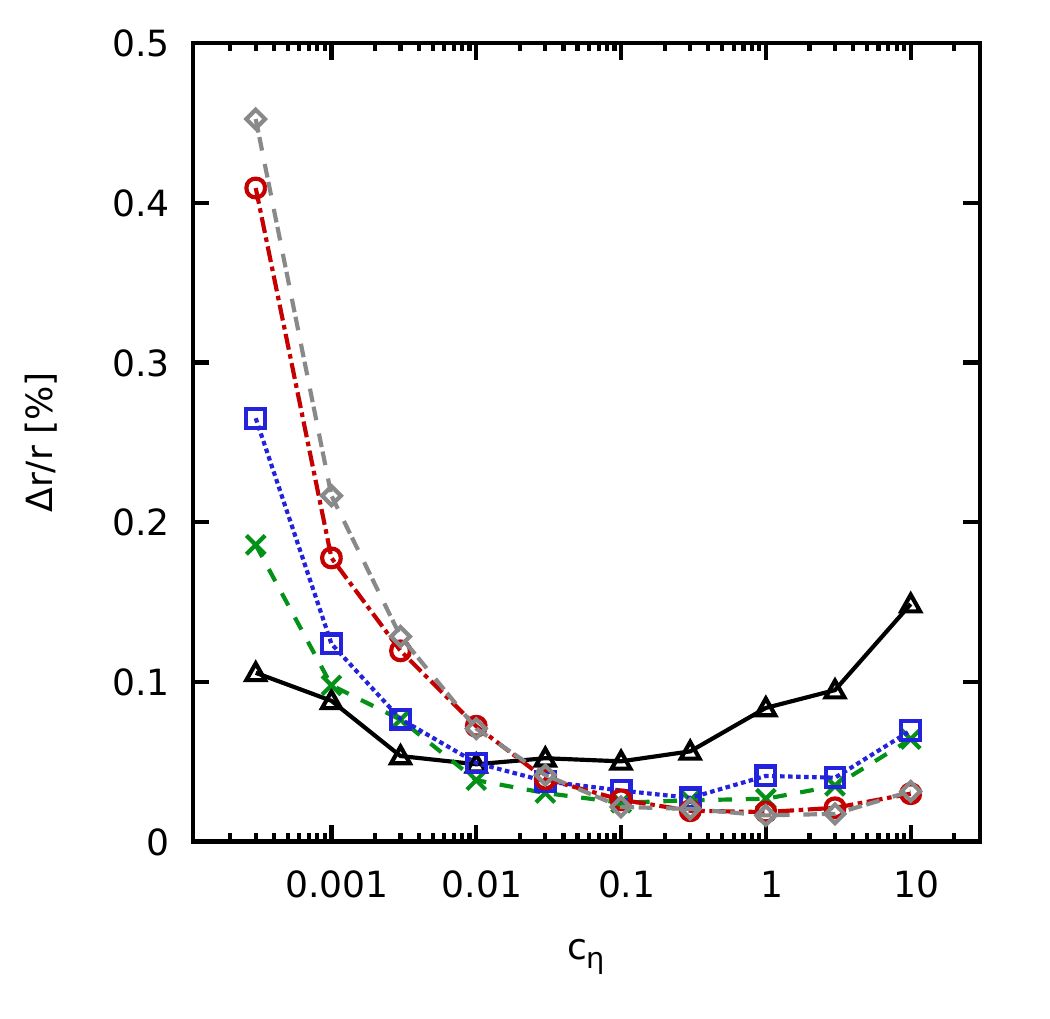}
\caption{\label{fig:UEG_N14_rs2_theta2_switch_probability} Left: PIMC acceptance probability to switch between two $\eta$-sectors for the UEG with $N=14$, $r_s=2$, $\Theta=2$ and for different $\eta$-transitions; center: estimated ratio of the partition functions [Eq.~(\ref{eq:r2})]; right: corresponding relative statistical uncertainty for equal total computation time.
}
\end{figure*}

\subsection{Fermionic free energy\label{sec:final}}

The final result for the free energy is given by~\cite{dornheim2024directfreeenergycalculation}
\begin{eqnarray}\label{eq:final}
    F_F = F_{B,0} - \frac{1}{\beta}\left\{
\textnormal{log}(S) + \sum_{i=1}^{N_\eta}\textnormal{log}\left(\frac{r(\eta_i,\eta_{i+1})}{c_{\eta_i}}\right)
    \right\}\ ,
\end{eqnarray}
with $\eta_1\equiv1$ and $\eta_{{N_\eta}+1}\equiv0$.
In principle, it would be sufficient to limit ourselves to $N_\eta=1$, i.e., to switch between the interacting and ideal limits in our simulation. In practice, however, the two configuration spaces might differ substantially. In that case, the acceptance probability for the switch update introduced in Sec.~\ref{sec:extended_ensemble} above would vanish, thereby threatening the ergodicity of the sampling scheme. With increasing $N_\eta$, the configuration spaces of adjacent $\eta$-sectors will always have sufficient overlap. This, moreover, comes at little additional computation cost as these simulations are exclusively carried out for the bosonic system that is sign-problem free.

The basic idea behind the $\eta$-ensemble approach for the direct PIMC based estimation of the free energy is sketched in Fig.~\ref{fig:scheme}. To estimate the free energy of the fermionic UEG, one has to start at a reference system for which the free energy is already known. Naively, one might start with the ideal Fermi gas (bottom left), which would require the evaluation of Eqs.~(\ref{eq:r}) and (\ref{eq:r2}) for the fermionic system. Such simulations would be subject to a massive sign problem, which is substantially more severe due to the high degree of quantum degeneracy in the ideal Fermi limit. While formally exact, this route (left vertical arrow) would be unfeasible except in the limit of high temperatures or few particles. To avoid this stifling limitation, we instead propose to start with the ideal Bose gas (bottom right). Since bosonic systems are sign-problem free, connecting the ideal and interacting Bose systems (right vertical error) is computationally cheap even if one has to introduce a large number of intermediate steps $N_\eta$. The final connection is then between the interacting Bose and Fermi systems [Eq.~(\ref{eq:F_and_S})], facilitated by the average sign $S$. The latter is an indispensable standard observable in direct PIMC simulations of Fermi systems~\cite{dornheim_sign_problem}. Moreover, it is evaluated in the interacting limit, where the sign problem is least severe. As a consequence, our scheme works with negligible computational overhead in all situations where fermionic PIMC simulations are feasible.
\begin{figure*}\centering
\includegraphics[width=0.74\textwidth]{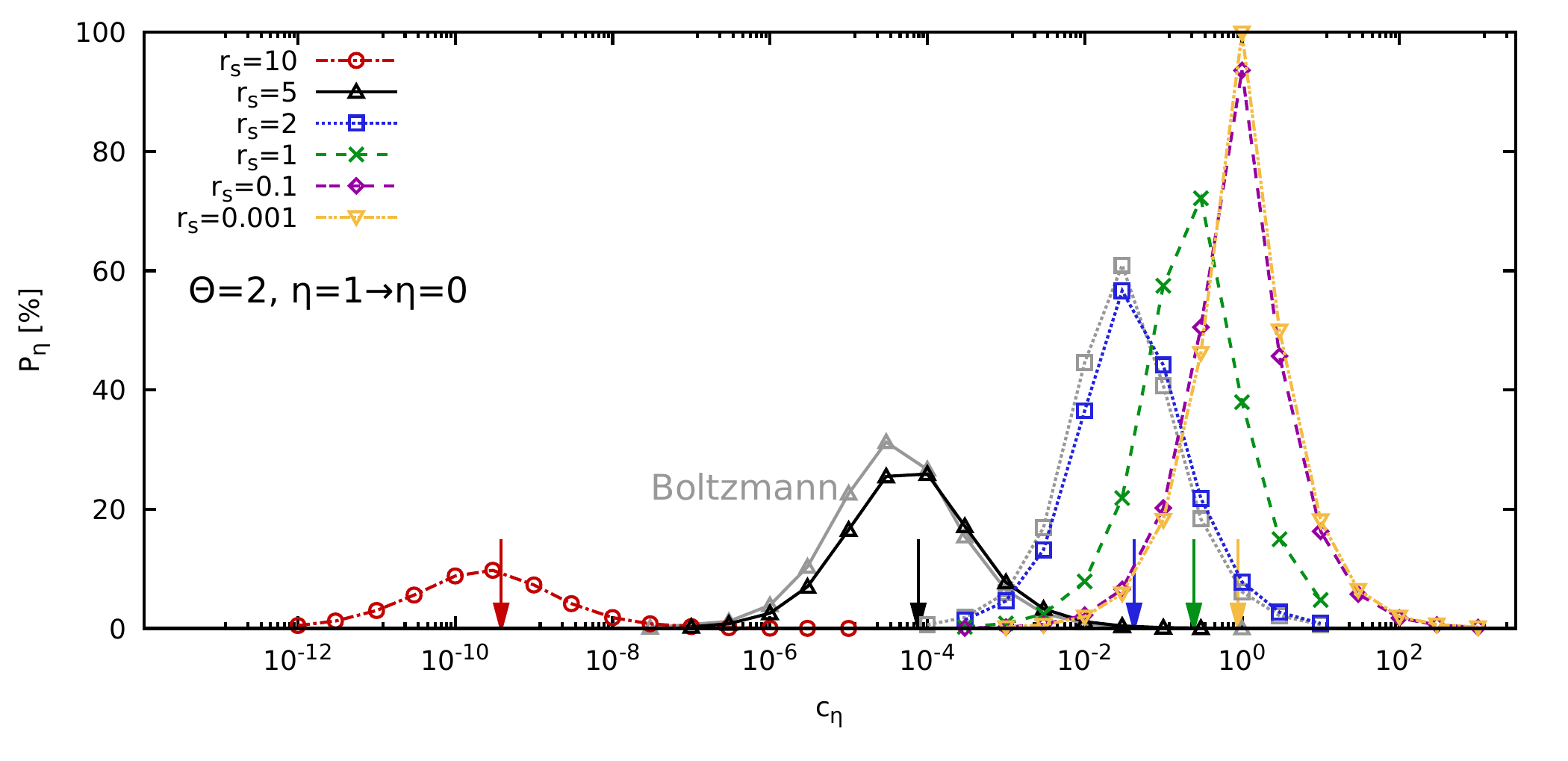}\\
\includegraphics[width=0.37\textwidth]{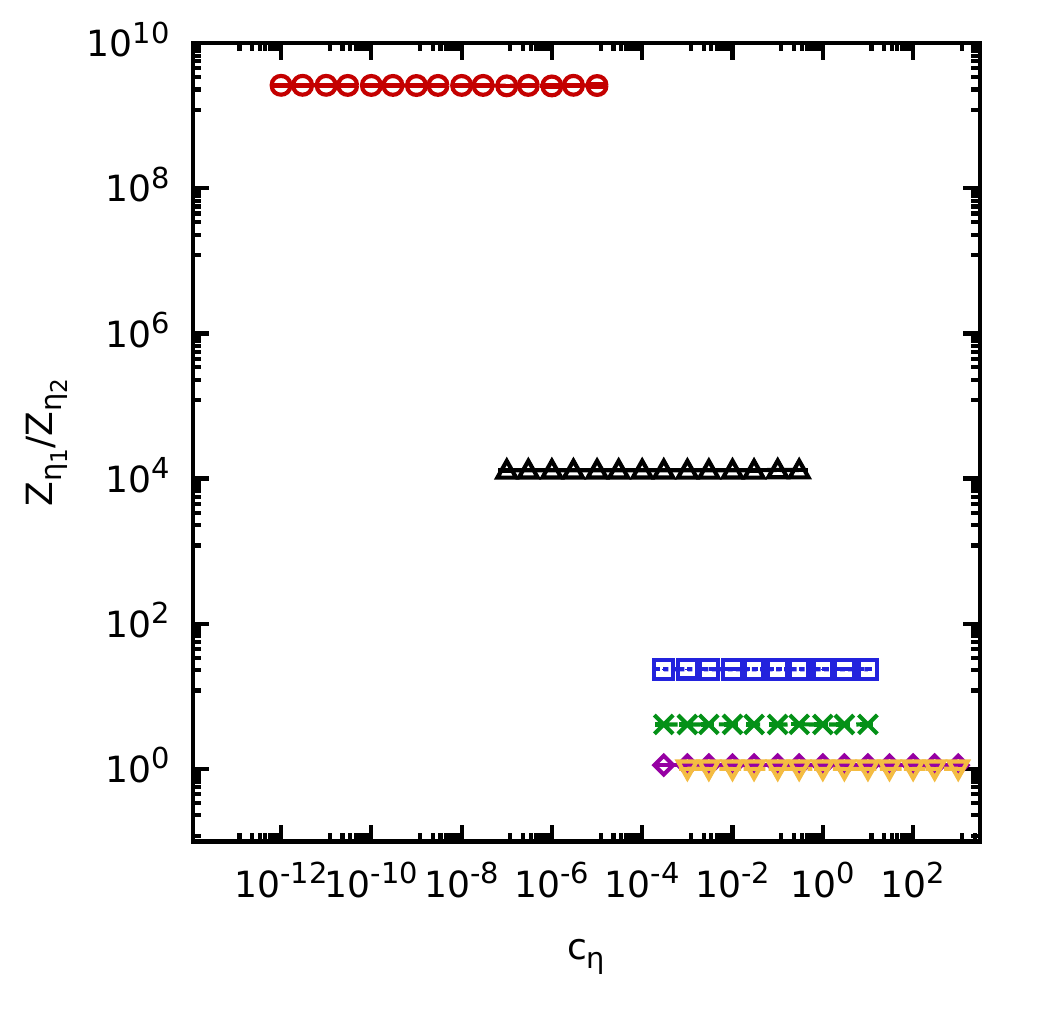}\includegraphics[width=0.37\textwidth]{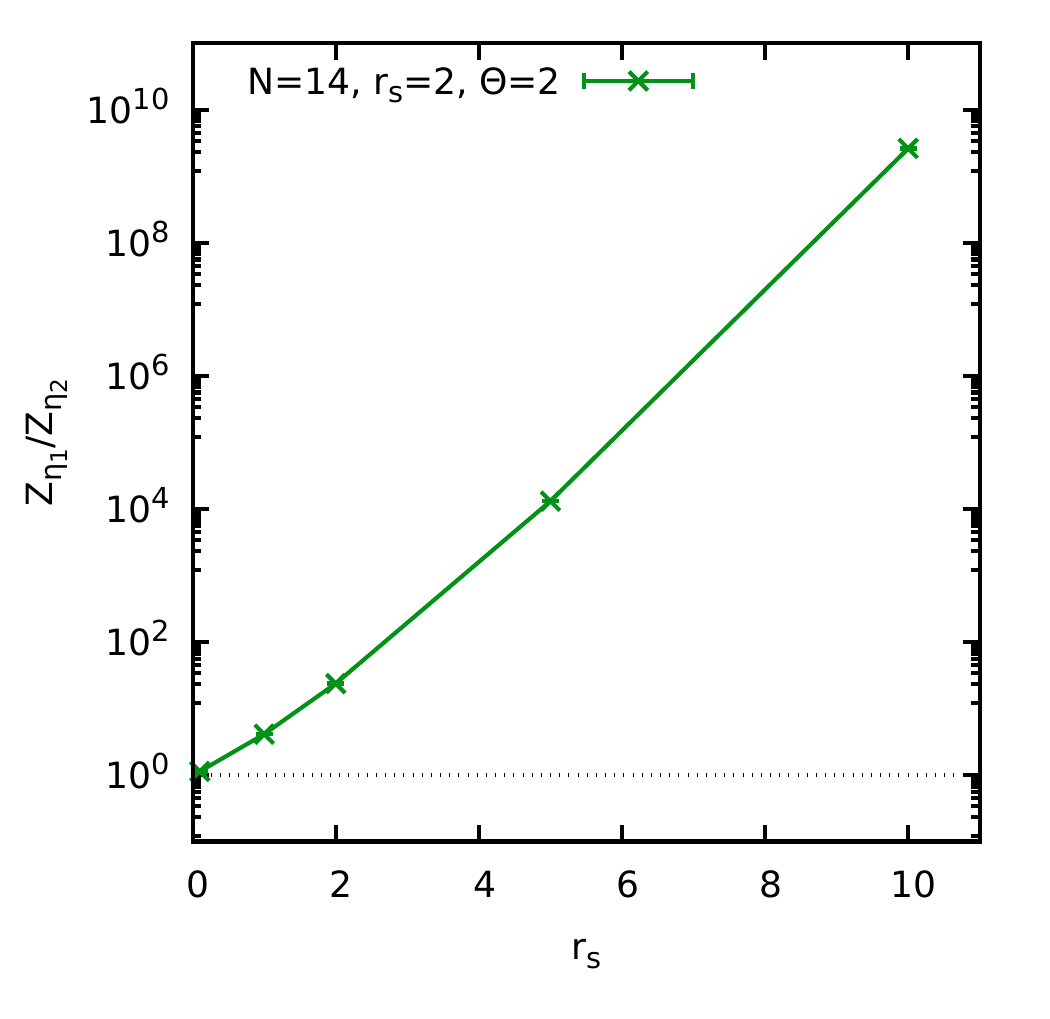}
\caption{\label{fig:UEG_N14_rs_theta2_switch_probability} Top: PIMC acceptance probability to switch between $\eta_1=1$ and $\eta_2=0$; bottom: corresponding ratio of the partition functions. Results have been obtained for the UEG with $N=14$, $\Theta=2$ and different $r_s$.
}
\end{figure*} 

We further note that Eq.~(\ref{eq:final}) directly outlines a decomposition of the full free energy into three distinct contributions:
\begin{eqnarray}\label{eq:decomposition}
    F_F = F_{B,0} + \Delta F_{B0,B} + \Delta F_{F,B}\ .
\end{eqnarray}
The first contribution is the free energy of the ideal Bose gas~\cite{dornheim2024directfreeenergycalculation,PhysRevResearch.2.043206,Zhou_2018}, which we compute from the corresponding partition function~\cite{Zhou_2018}
\begin{eqnarray}\label{eq:Z_ideal_Bose_Fermi}
    Z_{\textnormal{B},0} = \left[\frac{1}{N^\uparrow!} \textnormal{det}\left( \mathbf{Z}_{\textnormal{B},0}\right)\right]^2\ ,
\end{eqnarray}
with the elements of the $N^\uparrow\times N^\uparrow$ matrix $\mathbf{Z}_{\textnormal{B},0}$ defined as
\begin{eqnarray}\label{eq:get_off_my_case}
    \mathbf{Z}_{\textnormal{B},0}(i,j) = \begin{cases}
    -i, & \text{if } i=j+1\\
    Z[\Omega,1/\Omega,\beta (1+j-i)], & \text{if } i \leq j\\
    0,              & \text{otherwise}\ ,
\end{cases}
\end{eqnarray}
where $i,j=0,\dots,N^\uparrow-1$. In essence, Eq.~(\ref{eq:Z_ideal_Bose_Fermi}) constitutes a compact form of the well-known recursion relation~\cite{krauth2006statistical}, and it only requires the single-particle partition function $Z[\Omega,1/\Omega,\beta (1+j-i)]$ at different temperatures, which is easy to evaluate from the usual plain-wave representation. For completeness, we note that the non-interacting free-energy $F_{\textnormal{F},0}$ can be computed analogously from the matrix $\mathbf{Z}_{\textnormal{F},0}$ with elements $\mathbf{Z}_{\textnormal{F},0}(i,j)$, where the $-i$ entries are replaced by $i$ in the first line of Eq.~(\ref{eq:get_off_my_case})~\cite{Zhou_2018}.

The second term in Eq.~(\ref{eq:decomposition}) is given by the interaction correction from the bosonic system,
    \begin{eqnarray}\label{eq:correction_eta}
    \Delta F_{B0,B} = -\frac{1}{\beta} \sum_{i=1}^{N_\eta}\textnormal{log}\left(\frac{r(\eta_i,\eta_{i+1})}{c_{\eta_i}}\right)\ .
\end{eqnarray}
The final term is purely quantum statistical in origin and connects the interacting Bose and Fermi systems:
\begin{eqnarray}\label{eq:correction_statistics}
    \Delta F_{F,B} =- \frac{1}{\beta}\textnormal{log}(S) \ .
    \end{eqnarray}
The respective relative importance of these three contributions depends on the physical regime of interest and is investigated based on our new PIMC simulations below.

\section{Results\label{sec:results}}

All the PIMC results that are discussed in the present work have been obtained using the open-source \texttt{ISHTAR} code~\cite{ISHTAR}, which is based on the extended ensemble sampling algorithm introduced in Ref.~\cite{Dornheim_PRB_nk_2021}.

\subsection{Algorithmic efficiency: general trends\label{sec:efficiency}}

Let us start our analysis of the PIMC approach to the free energy of the UEG with an investigation of the algorithmic parameters and efficiency. In Fig.~\ref{fig:UEG_N14_rs2_theta2_switch_probability}, we show results for $N=14$, $r_s=2$, and $\Theta=2$. The left panel shows the acceptance probability of the Metropolis update that switches between the $\eta$-sectors as a function of the free parameter $c_\eta \equiv c_{\eta_1}$. 
The black triangles show results for $\eta_1=1$ and $\eta_2=0$, i.e., for a single transition between the interacting and ideal limits and we find a strong dependence on $c_\eta$ with a maximum acceptance probability of around $60\%$ for $c_\eta\sim0.05$. In the central panel, we show corresponding results for the ratio of partition functions, cf.~Eq.~(\ref{eq:r2}); as it is expected, the PIMC results do not depend on $c_\eta$, which is a useful cross-check of our implementation. The right panel shows the associated statistical uncertainty for equal total computation time. For small and large $c_\eta$, the acceptance probability of the transition update is small. Consequently, transitions between the $\eta$-sectors occur only infrequently and adjacent measurements in the Markov chain are strongly correlated. As a consequence, the relative statistical error increases in both limits. Conversely, we find a (flat) minimum around $c_\eta\sim0.05$, i.e., around the maximum in the transition probability.

We next return to the left panel and consider the green crosses and blue squares, which correspond to PIMC simulations with $\eta_1=1$ and $\eta_2=0.5$ as well as $\eta_1=0.5$ and $\eta_2=0$, respectively. In other words, we have introduced an intermediate step between the interacting and ideal limits by setting $N_\eta=2$ in Eq.~(\ref{eq:final}).
First, we find that the maximum in the acceptance probability of the transition update is increased and attains a value of $\sim70\%$ for both curves. This is expected as the respective configuration spaces for $\eta_1$ and $\eta_2$ become increasingly similar for larger $N_\eta$. Second, the position of the maximum is shifted to larger $c_\eta$ for the same reason. In particular, the potential energy of the UEG is negative due to the interaction between the electrons and the positive uniform background~\cite{review}. From the expression for the acceptance probability of the transition update in Eqs.~(\ref{eq:A1}) and (\ref{eq:A2}), we thus immediately see that the larger $\eta_1$ is favored over the smaller $\eta_2$. The free parameter $c_\eta$ has been introduced to counteract this imbalance, which, in practice, means that it should decrease the relative weight of the $\eta_1$-sector. Finally, the red circles and gray diamonds show the respective acceptance probability for $\eta_1=1$ and $\eta_2=0.9$ as well as for $\eta_1=0.1$ and $\eta_2=0$, i.e., for two steps in a set-up with $N_\eta=10$. Here, the differences between the simulated configuration spaces is so small that we find a maximum in the acceptance probability of $>80\%$ around $c_\eta=1$.

Moving on to the central panel, we confirm our earlier point that the ratio of the corresponding partition functions is independent of $c_\eta$ in all cases. While expected, it is still remarkable that we obtain such a level of accuracy even for simulations with an acceptance probability of $P\ll1\%$. In practice, this is reflected by the statistical uncertainty that is shown in the right panel. While the error bar is relatively flat for the black triangles, we find a sharp increase in the other curves for decreasing $c_\eta$ where the transition probability drops below $1\%$. While asymptotically correct, these simulations are still very inefficient due to strong correlations between adjacent elements in the Markov chain. For the PIMC simulations with $N_\eta=2$ (green crosses and blue squares), we find errors of approximately half the size compared to $N_\eta=1$, which means that the overall efficiency remains unchanged; this is a consequence of the central limit theorem that gives a scaling $\Delta O\sim 1/\sqrt{N_\textnormal{MC}}$ for the Monte Carlo error of any observable $O$ with the number of Monte Carlo samples $N_\textnormal{MC}$.
The error bars are further decreased for $N_\eta=10$, but this seeming advantage is overcompensated by the larger number of Monte Carlo simulations needed to capture the full transition between $\eta=1$ and $\eta=0$, cf.~Eq.~(\ref{eq:final}). We emphasize again that all simulations shown in Fig.~\ref{fig:UEG_N14_rs2_theta2_switch_probability} have been carried out using the same amount of compute time.


\begin{figure}\centering
\includegraphics[width=0.37\textwidth]{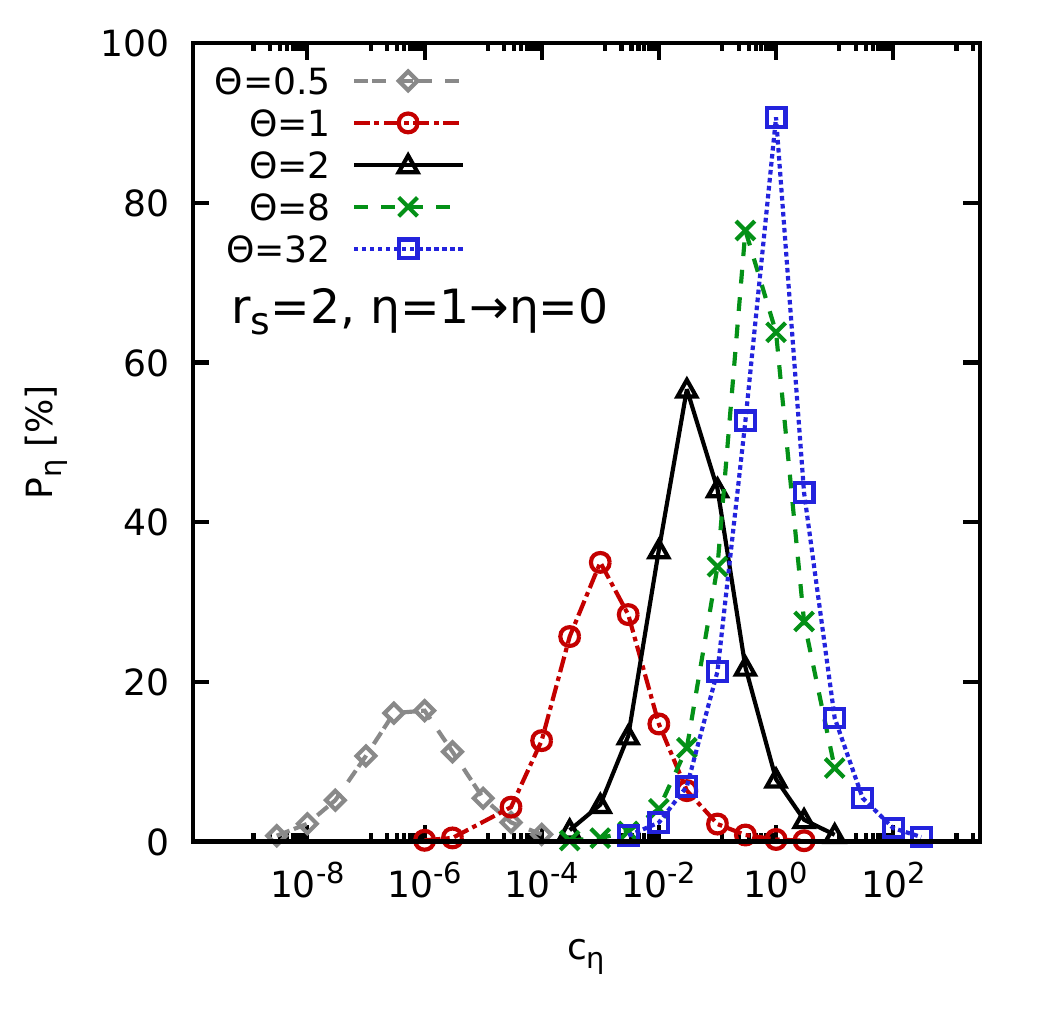}\\
\includegraphics[width=0.37\textwidth]{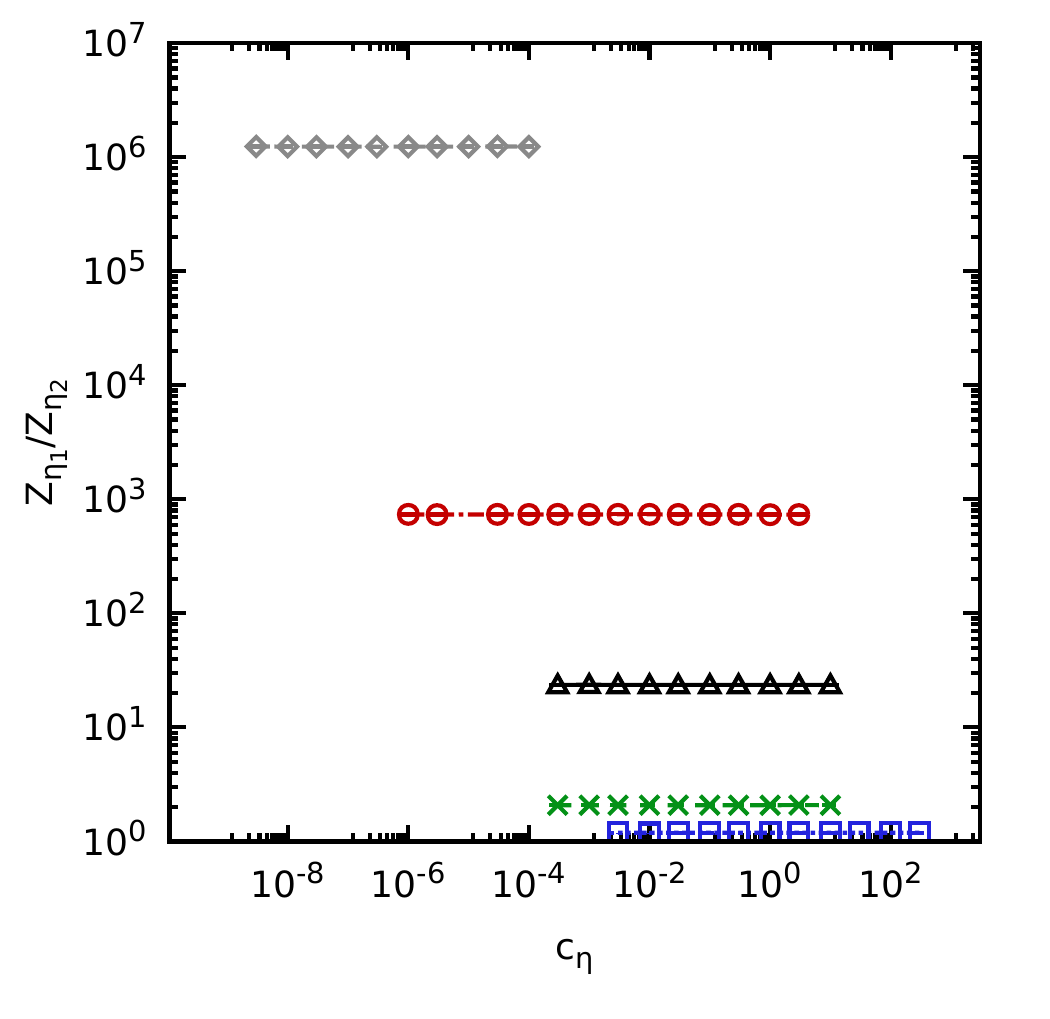}
\caption{\label{fig:UEG_N14_rs2_theta_switch_probability} Top: PIMC acceptance probability to switch between $\eta_1=1$ and $\eta_2=0$; bottom: corresponding ratio of the partition functions. Results have been obtained for the UEG with $N=14$, $r_s=2$ and different $\Theta$.
}
\end{figure} 

\begin{figure}\centering
\includegraphics[width=0.37\textwidth]{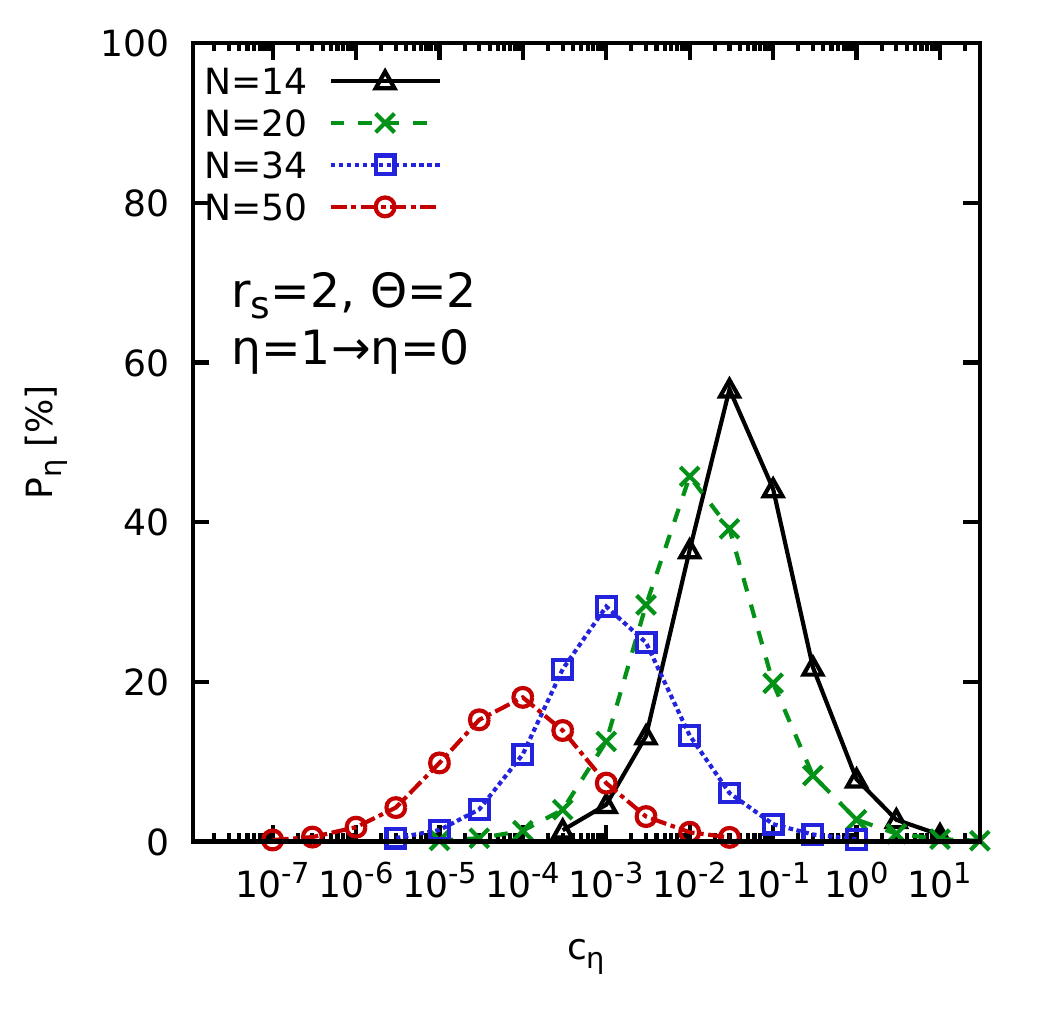}\\
\includegraphics[width=0.37\textwidth]{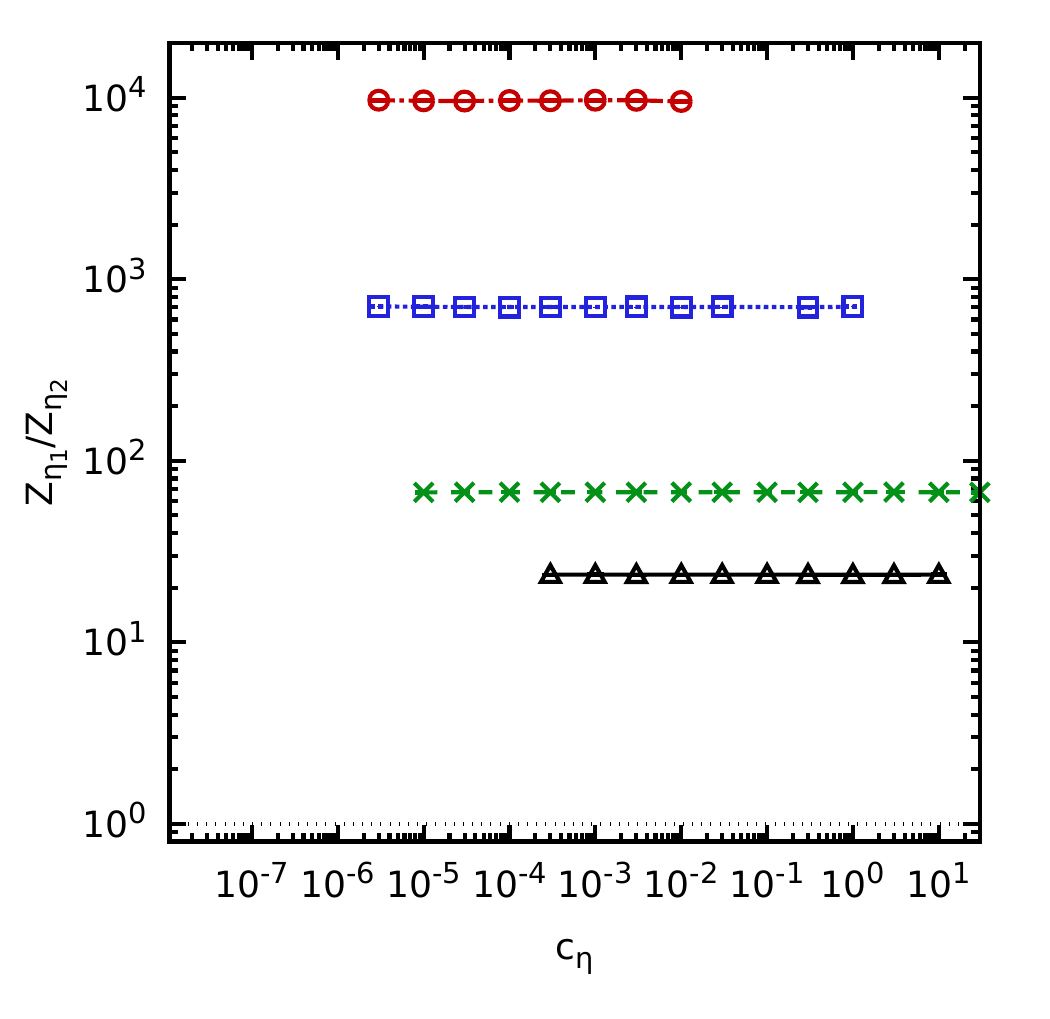}
\caption{\label{fig:UEG_N_rs2_theta2_switch_probability} Top: PIMC acceptance probability to switch between $\eta_1=1$ and $\eta_2=0$; bottom: corresponding ratio of the partition functions. Results have been obtained for the UEG with $\Theta=2$, $r_s=2$ and different $N$.
}
\end{figure} 

In the top panel of Fig.~\ref{fig:UEG_N14_rs_theta2_switch_probability}, we show the dependence of the acceptance probability of the $\eta$-transition update on $c_\eta$ for $N_\eta=1$ for various values of the density parameter spanning four orders of magnitude. For $r_s=0.001$ (yellow downward triangles and $r_s=0.1$ (purple diamonds), the system is only weakly interacting. Consequently, the configuration spaces between $\eta_1=1$ and $\eta_2=0$ are nearly identical, and we find a pronounced maximum around $c_\eta=1$. For $r_s=0.001$, the acceptance probability nearly equals $100\%$, which corresponds to the limit of the ideal Fermi gas where both $\eta$-sectors would be identical.

Upon decreasing the density, Coulomb coupling effects become more important, which manifests in two ways. First, the dissimilarity of the two simulated $\eta$-sectors leads to an overall decrease of the maximum possible update acceptance probability, which barely reaches $10\%$ for $r_s=10$ (red circles). Second, the position of the maximum gets shifted to smaller $c_\eta$; this shift is given by more than nine orders of magnitude for $r_s=10$.
In the bottom left panel of Fig.~\ref{fig:UEG_N14_rs_theta2_switch_probability}, we show the corresponding results for the ratio of the two partition functions; as usual, it is independent of $c_\eta$ in all cases, and its dependence on $r_s$ is shown in the bottom right panel of the same figure. Indeed, the approximately exponential increase of this ratio with $r_s$ is equivalent to the shift in the maxima in the top panel, where the vertical coloured arrows indicate the reverse ratio. In other words, the optimal choice of $c_\eta$ would exactly compensate the imbalance of the two partition functions, giving the two $\eta$-sectors the same weight in our simulations.

A final interesting consideration concerns the origin in the disparity between the two configuration spaces. On the one hand, one might assume that it is a direct interaction effect that leads to a net decrease in all energies and, hence, to a larger partition function compared to the ideal reference system. A related, though more subtle possibility is given by the effect of the Coulomb coupling onto the permutation structure of the system. In particular, Coulomb coupling effects are known to reduce the formation of permutation cycles~\cite{Dornheim_permutation_cycles}, which might be an alternative source for the disparity between the limits of $\eta_1$ and $\eta_2$.
To resolve the respective impact of these effects, we have carried out additional PIMC simulations for distinguishable quantum particles (i.e., boltzmannons), and the results are included as the grey symbols in the top panel of Fig.~\ref{fig:UEG_N14_rs_theta2_switch_probability}. Clearly, the boltzmannonic results closely resemble the bosonic data, thereby ruling out any disparities in the permutation structure as the main cause for the observed trends.
If anything, the latter somewhat counteracts the direct interaction effect as the finite weight of additional permutation sectors somewhat increases the partition function in the $\eta=0$ limit compared to the interacting case, although this effect is small in practice.

Let us next consider the impact of the temperature, which is investigated in detail in Fig.~\ref{fig:UEG_N14_rs2_theta_switch_probability} for the UEG with $N=14$ and $r_s=2$. 
Overall, we find a similar qualitative picture as for Fig.~\ref{fig:UEG_N14_rs_theta2_switch_probability} discussed above: with increasing temperature, the $\eta_1=1$ limit becomes more ideal and we get overall larger acceptance probabilities for $c_\eta$ closer to unity. Conversely, interaction effects play a more important role for lower temperatures, with the, by now, expected results. Note that we chose to not go below half the Fermi temperature here since the average sign $S$, which is an indispensable ingredient to the total free energy of the UEG [cf.~Eq.~(\ref{eq:final})] cannot be resolved in this regime. The bottom panel of Fig.~\ref{fig:UEG_N14_rs2_theta_switch_probability}
shows the corresponding ratios of the partition functions, again with the expected trends.

The final system parameter to be investigated here is given by the dependence on the system size, which is shown in Fig.~\ref{fig:UEG_N_rs2_theta2_switch_probability}. Evidently, increasing the number of particles leads to similar trends compared to reducing the density or the temperature. This is, at a first glance, rather surprising as the coupling strength is generally independent of the particular system size. Indeed, the likely origin of this trend is more subtle. In the noninteracting limit ($\eta_2=0$), there is no penalty for two particles coming very close to each other. If, starting from such a configuration, a switch to the interacting limit of $\eta_1=1$ is proposed, the Coulomb repulsion would lead to a penalty that increases exponentially with decreasing distance $r$. With increasing system size, there is always a higher probability for the occurrence of such particle pair configurations, which lead to a supra-linear penalty in the transition probability. As a consequence, transitions between the $\eta_1$ and $\eta_2$ sectors become increasingly unlikely for large systems, which, in particular, explains the reduced maxima in $P_\eta$.
Therefore, the evaluation of the free energy for larger systems generally requires a larger number of intermediate steps $N_\eta$.

\subsection{Algorithmic efficiency: electron liquid\label{sec:el}}

\begin{figure}\centering
\includegraphics[width=0.37\textwidth]{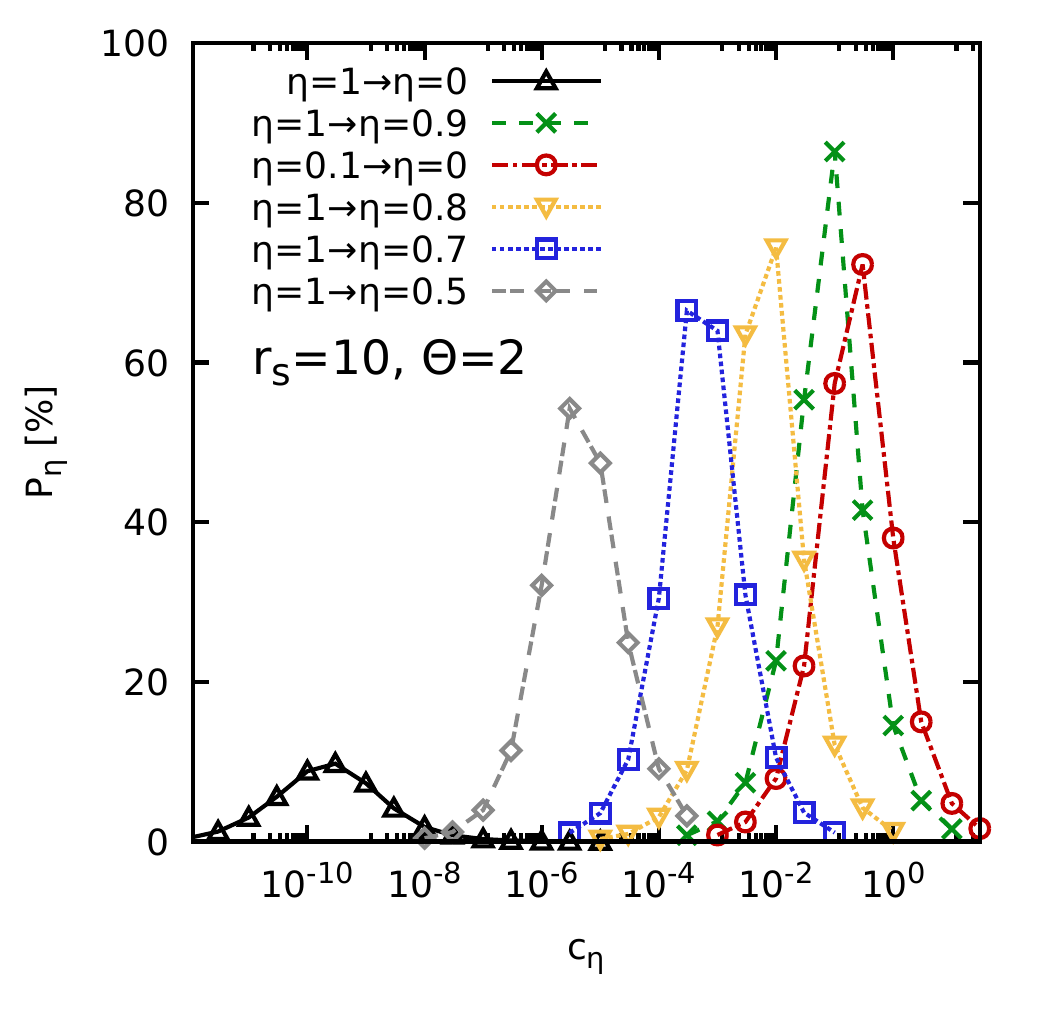}
\caption{\label{fig:UEG_N_rs10_theta2_switch_probability} Acceptance probability for the transition between numerous different $\eta$-sectors as a function of the free parameter $c_\eta$ for the UEG with $N=14$, $r_s=10$, and $\Theta=2$. 
}
\end{figure} 

\begin{figure*}\centering
\includegraphics[width=0.74\textwidth]{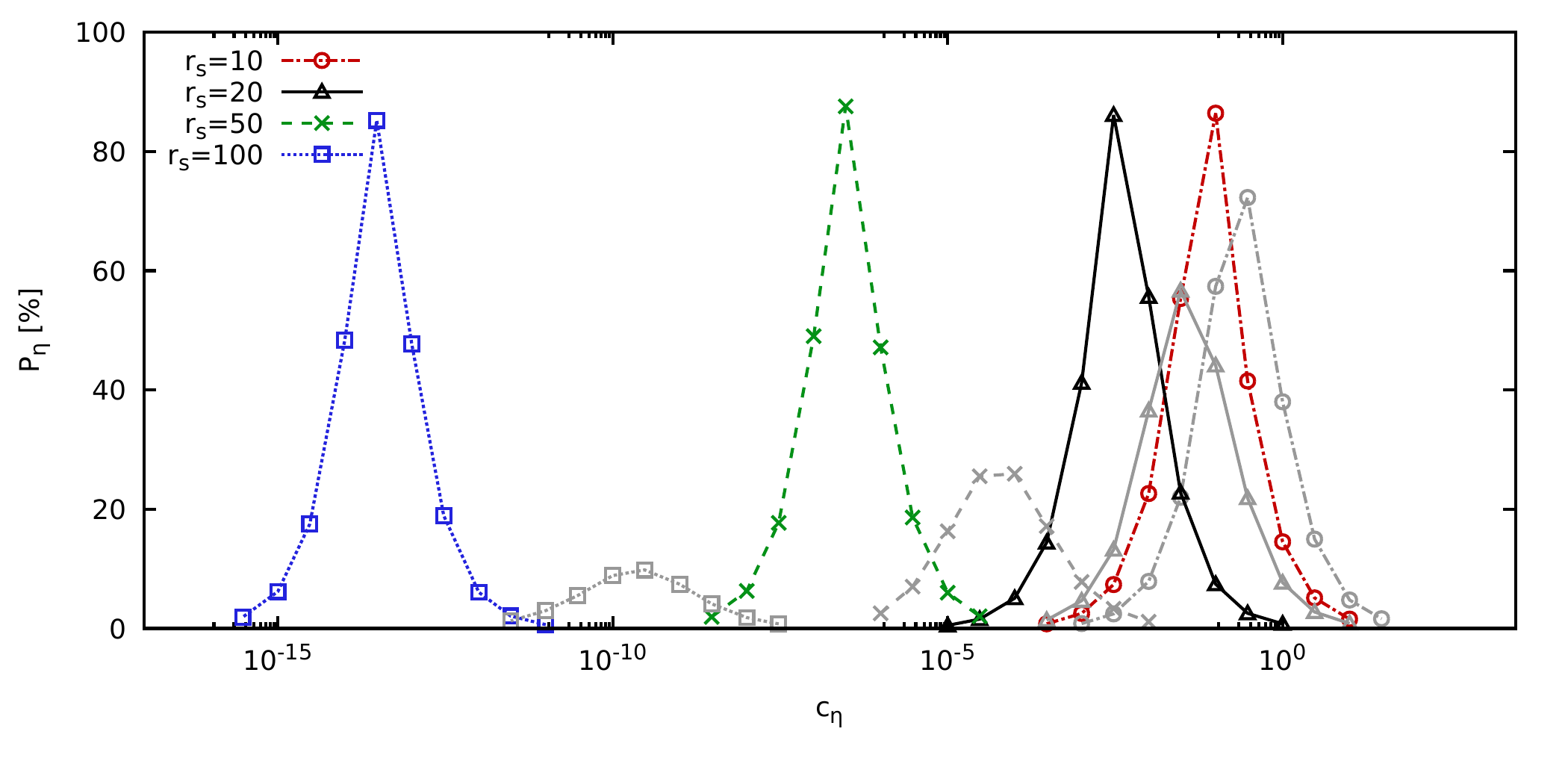}
\caption{\label{fig:Switch_Electron_Liquid} Colored (grey): acceptance probability for the transition between $\eta$-sectors with $\eta_1=1$ and $\eta_2=0.9$ ($\eta_1=0.1$ and $\eta_2=0$) for the UEG with $N=14$, $\Theta=2$, and various values of the density parameter $r_s$ in the strongly coupled electron liquid regime as a function of the free algorithmic parameter $c_\eta$.
}
\end{figure*} 

\begin{figure*}\centering
\includegraphics[width=0.34\textwidth]{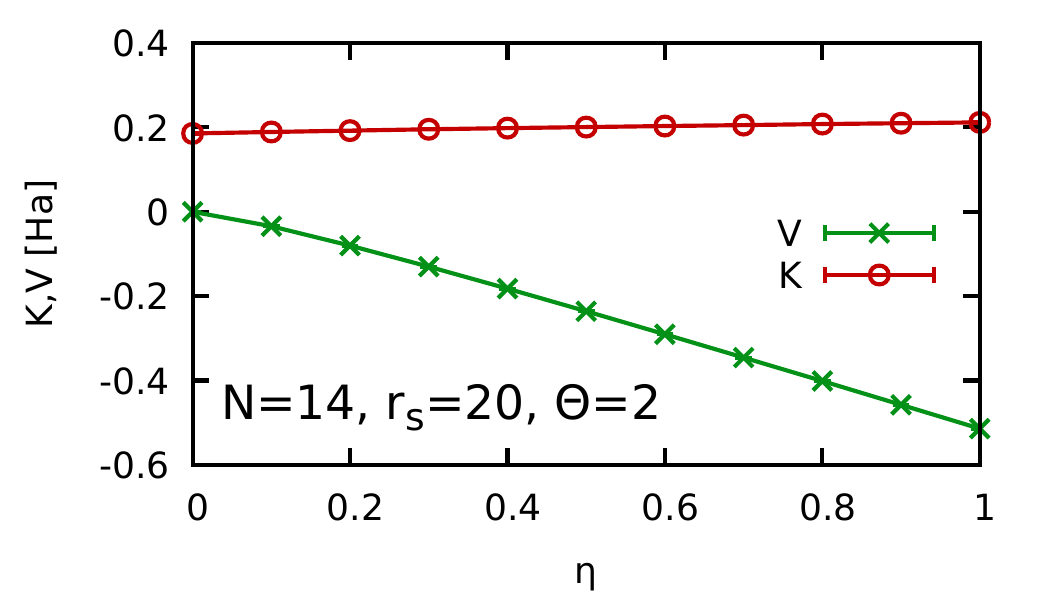}\hspace*{-0.3cm}
\includegraphics[width=0.34\textwidth]{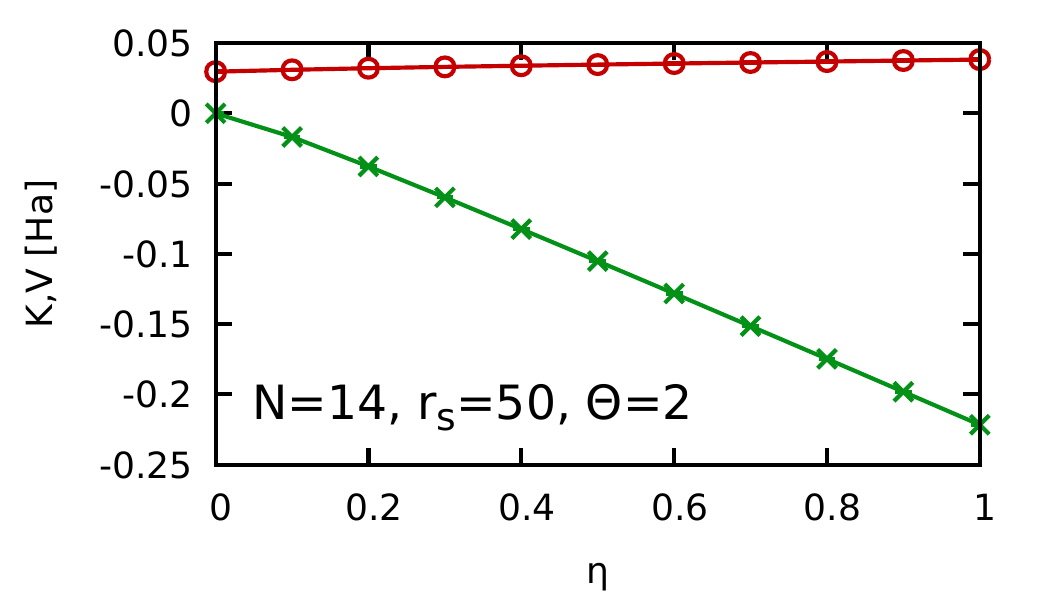}\hspace*{-0.3cm}
\includegraphics[width=0.34\textwidth]{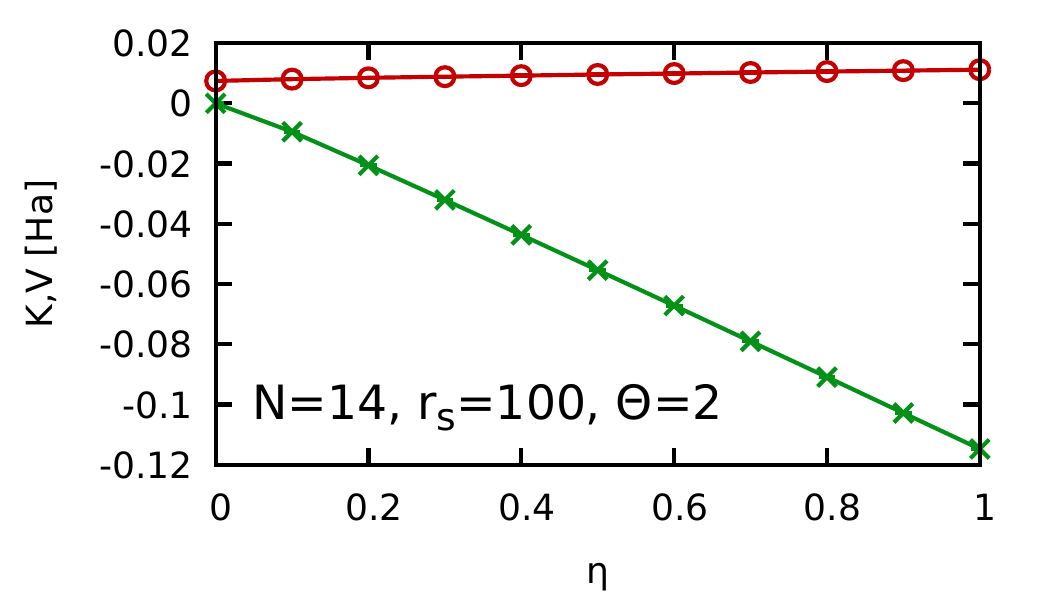}
\caption{\label{fig:single} Dependence of the kinetic energy (red) and interaction energy (green) on $\eta$ for the UEG with $N=14$ and $\Theta=2$ for $r_s=20$ (left), $r_s=50$ (center), and $r_s=100$ (right).
}
\end{figure*} 

From Sec.~\ref{sec:efficiency}, it has become clear that the estimation of the free energy is more involved in the strongly coupled electron liquid regime due to the increasing disparity between the ideal and interacting limits. In practice, this obstacle can be easily removed by increasing the number of intermediate steps $N_\eta$, which is investigated in more detail in the following.

In Fig.~\ref{fig:UEG_N_rs10_theta2_switch_probability}, we illustrate the acceptance probability of the $\eta$-transition update for $r_s=10$ and $\Theta=2$ for different values of $N_\eta$ and different transitions; the black upwards-triangles show results for $N_\eta=1$ (as shown in Fig.~\ref{fig:UEG_N14_rs_theta2_switch_probability} above) and have been included as a reference. Increasing the number of transition steps to $N_\eta=2$ in Eq.~(\ref{eq:final}) leads to the grey diamonds ($\eta_1=1$ and $\eta_2=0.5$), which are characterized by a maximum acceptance probability exceeding $50\%$, i.e., with a greatly increased ergodicity compared to $N_\eta=1$. Further reduction of the step size to $\Delta\eta=\eta_1-\eta_2=0.3$ (blue squares) and $\Delta\eta=0.2$ (yellow downwards triangles) somewhat further increases the sampling efficiency, without being decisive. Finally, the green crosses and red circles both correspond to $N_\eta=10$, but to the opposite ends of the full $\eta$-range. 
Interestingly, the two curves differ significantly both in the position and height of their respective maximum. 

This effect is further exacerbated in the case of strong Coulomb coupling, as it can be seen in Fig.~\ref{fig:Switch_Electron_Liquid}. To be more specific, the colored symbols show results for $\eta_1=1$ and $\eta_2=0.9$ for $r_s=10$ (red circles), $r_s=20$ (black triangles), $r_s=50$ (green crosses), and $r_s=100$ (blue squares). In addition, the corresponding grey symbols show results for the same conditions but $\eta_1=0.1$ and $\eta_2=0$. Evidently, there appears a widening gap between the different $\eta$-sectors for lower densities and stronger coupling. Interestingly, the maxima of the acceptance probability for the $\eta$-transition update seem to be independent of $r_s$ for $\eta_1=1$ and $\eta_2=0.9$, whereas the position of the maximum is shifted to smaller $c_\eta$. This implies that the two $\eta$-sectors are fairly similar (independent of the density), but there appears a constant shift in energy between $\eta_1$ and $\eta_2$ that has to be balanced by an appropriate choice of $c_\eta$. This is unsurprising in so far as configurations with two overlapping particles are almost equally prohibited in both sectors.
In contrast, the two sectors are separated by a lower shift in (free) energy for $\eta_1=0.1$ and $\eta_2=0$, hence the position of the maximum of the acceptance probability of the transition is located at larger $c_\eta$. At the same time, the maximum acceptance is drastically reduced due to the increasing dissimilarity of the two configuration spaces.

\begin{figure}\centering
\includegraphics[width=0.37\textwidth]{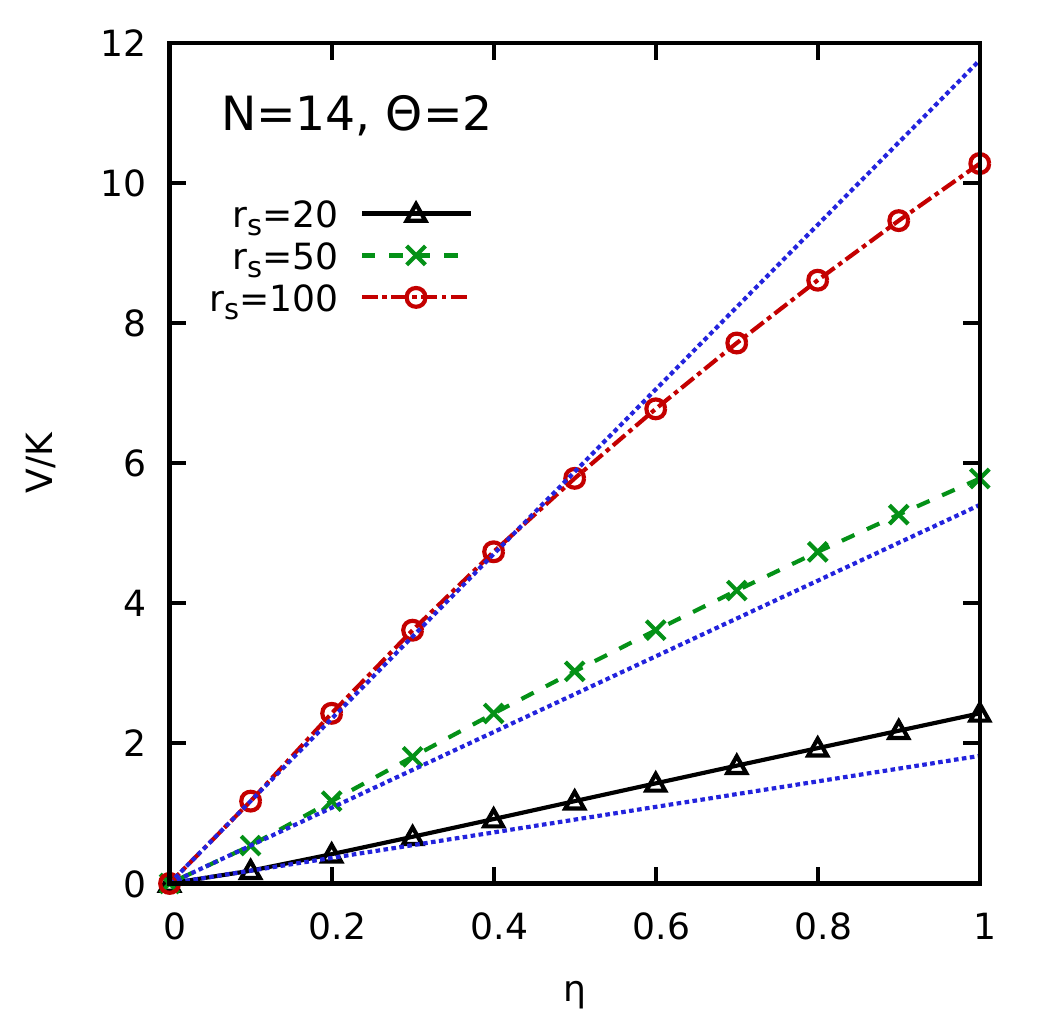}
\caption{\label{fig:Kopplung} Ratio of the potential to kinetic energy as a function of the free parameter $\eta$ for the UEG with $N=14$ and $\Theta=2$ for $r_s=20$ (black triangles), $r_s=50$ (green crosses), and $r_s=100$ (red circles). The dotted blue lines are linear interpolations for $\eta\leq0.1$ that have been included as a guide-to-the-eye.
}
\end{figure} 

\begin{figure}\centering
\includegraphics[width=0.37\textwidth]{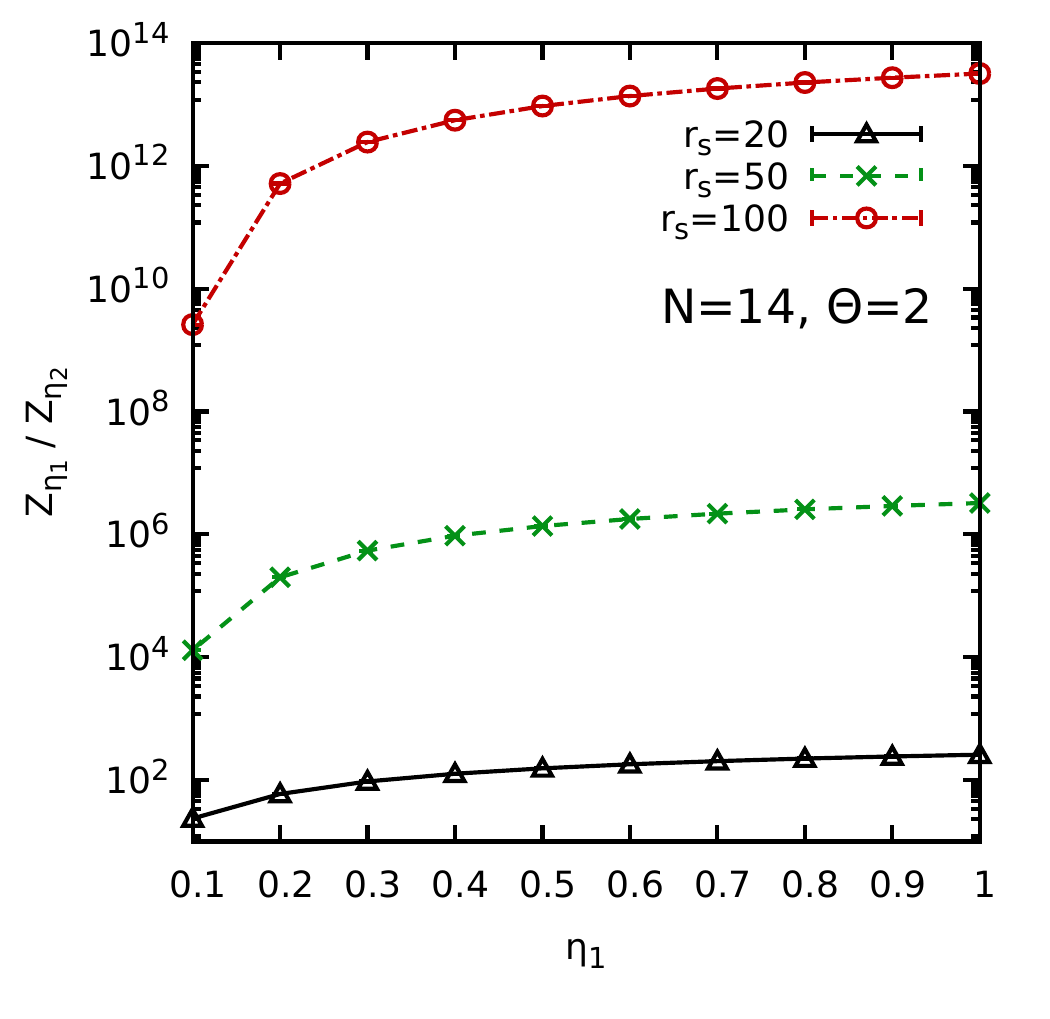}
\caption{\label{fig:eta_dependence} Ratio of partition functions as a function of $\eta_1$ for the UEG with $N=14$, $\Theta=2$ and $N_\eta=10$ for $r_s=20$ (black triangles), $r_s=50$ (green crosses), and $r_s=100$ (red circles).
}
\end{figure}

\begin{figure*}\centering
\includegraphics[width=0.74\textwidth]{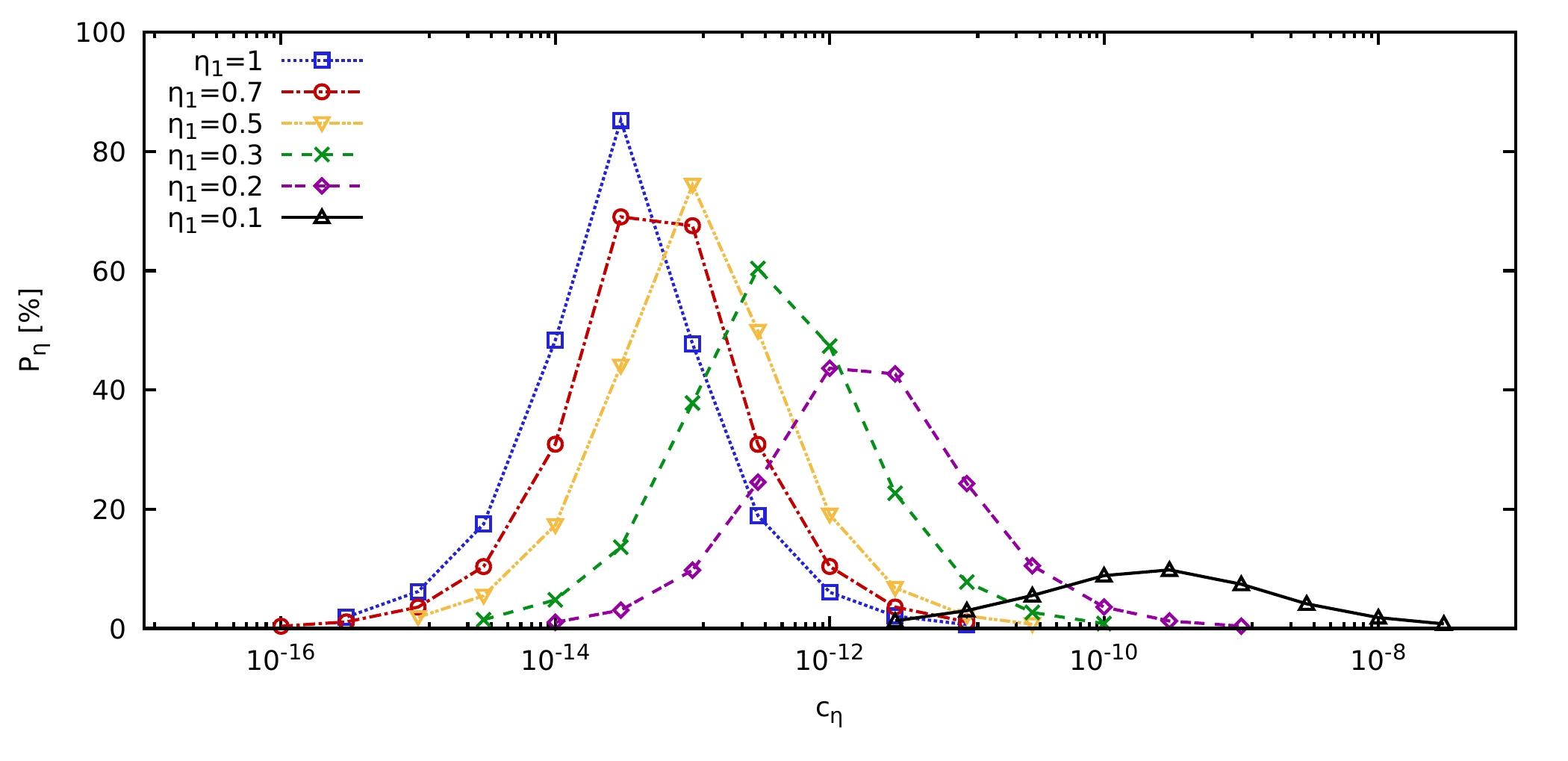}
\caption{\label{fig:Switch_N14_rs100} Acceptance probability for the transition between $\eta$-sectors for the UEG with $N=14$, $r_s=100$, and $\Theta=2$.
}
\end{figure*}

In Fig.~\ref{fig:single}, we show PIMC results for the kinetic (red circles) and potential (green crosses) energy for all three values of the density parameter as a function of $\eta$.
Evidently, the kinetic energy is comparably unaffected by the scaling of the potential term, whereas the effects of $\eta$ are more direct on the latter. Fig.~\ref{fig:Kopplung} shows corresponding results for the coupling parameter $\Gamma=V/K$, i.e., the ratio of the two energy contributions. Naturally, it holds $\Gamma=0$ for $\eta=0$ by definition. This limit is followed by a non-trivial non-linear increase in all three cases, where the dotted blue lines merely serve as a guide-to-the-eye. At the same time, we find no direct connection between this energy dependence and the trends observed in Fig.~\ref{fig:Switch_Electron_Liquid}. Finally, Fig.~\ref{fig:eta_dependence} shows the corresponding dependence of the ratio of the partition functions on $\eta_1$, where the transition from $\eta_1=0.1$ to $\eta_2=0$ clearly occupies a special place. This is investigated in more detail for the case of $r_s=100$ in Fig.~\ref{fig:Switch_N14_rs100}, where we show the acceptance probability of the transition update as a function of $c_\eta$ for $N_\eta=10$ for a number of transitions; both the reduction in the maximum and the shift of its position to larger $c_\eta$ are monotonic and fit the respective ratio of partition functions depicted in Fig.~\ref{fig:eta_dependence}.

These observations lead us to the following practical remarks: (i) to ensure ergodicity, i.e., an appropriate sampling of the full configuration space defined by Eq.~(\ref{eq:Z_extended}), we aim for an acceptance probability of $P_\eta\gtrsim5\%$; (ii) for lower temperatures, larger systems, and increasing $r_s$, the number of $\eta$-steps for the evaluation of the full free energy [Eq.~(\ref{eq:final})] should be increased; (iii) the $\eta$-grid for the evaluation of Eq.~(\ref{eq:final}) does not have to be equidistant. It makes sense to use a finer grid towards the limit of small $\eta$ to ensure a sufficient acceptance probability everywhere; (iv) in general, the optimal choice for the free algorithmic parameter $c_\eta$ depends on $\eta$.

\subsection{Exchange--correlation free energy\label{sec:XC}}

\begin{table*}
\caption{\label{tab:UEG}PIMC results for the UEG finite-size corrected XC-free energy per particle $F_\textnormal{xc}/N + \Delta F_\textnormal{xc}$, the corresponding finite-size correction $\Delta F_\textnormal{xc}$ computed with the \texttt{UEGPY} code~\cite{uegpy}, the (non-size corrected) free energy per particle $F_\textnormal{F}/N$, the free energy of the noninteracting Bose reference system $F_{\textnormal{B},0}$, the average sign $S$ of the interacting UEG, and the average sign $S_0$ of the noninteracting Fermi gas. Our new results for the thermodynamic limit ($N\to\infty$, bottom segment) have been obtained from empirical linear extrapolations, see the main text. The error bars of the GDSMFB data~\cite{groth_prl} correspond to the nominal uncertainty of $\sim0.3\%$; note that the validity range of the GDSMFB parametrization is limited to $r_s\leq20$.}
\begin{ruledtabular}
\begin{tabular}{crlllll}
    & & $r_s=2$ & $r_s=10$ & $r_s=20$ & $r_s=50$ & $r_s=100$\\
    \colrule\\[-1ex]
 $N=14$ & $F_\textnormal{xc}/N+\Delta F_\textnormal{xc}$&$-0.18549(10)$ & $-0.055198(9)$ & $-0.031317(3)$ & $-0.0141708(4)$ & $-0.00756663(11)$\\ 
 & $\Delta F_\textnormal{xc}$& $0.057741$ & $0.004517$ & $0.001409$ & $0.000297$ & $0.000092$\\[+1ex]
 & $F_\textnormal{F}/N$& $-2.09576(6)$ & $-0.133816(9)$ & $-0.051251(2)$ & $-0.0174369(4)$ & $-0.00839964(11)$\\
 & $F_{\textnormal{B},0}$ & $-3.85356$ & $-0.15414$ & $-0.0385356$ & $-0.0061657$ & $-0.00154143$\\[+1ex]
 & $S$& $0.5529(2)$ & $0.8921(2)$ & $0.97097(11)$ & $0.99760(2)$ & $0.999800(5)$\\
 & $S_0$& $0.3234(2)$ & $\dots$ & $\dots$ & $\dots$ & $\dots$\\[+1ex]\colrule\\[-1ex]
 $N=20$ & $F_\textnormal{xc}/N+\Delta F_\textnormal{xc}$&$-0.18596(13)$ & $-0.055223(9)$ & $-0.031318(3)$ & $-0.0141759(4)$ & $-0.00756390(10)$\\ 
 & $\Delta F_\textnormal{xc}$& $0.045469$ & $0.003381$ & $0.001041$ & $0.000216$ & $0.000066$\\[+1ex]
 & $F_\textnormal{F}/N$& $-2.14122(7)$ & $-0.134996(8)$ & $-0.051457(2)$ & $-0.0063608$ & $-0.00839382(8)$\\ 
 & $F_{\textnormal{B},0}$ & $-3.97549$ & $-0.15902$ & $-0.0397549$ & $-0.0174425(4)$ & $-0.00159020$\\[+1ex]
 & $S$& $0.4174(2)$ & $0.8463(6)$ & $0.9577(3)$ & $0.99657(3)$ & $0.999750(7)$\\
 & $S_0$& $0.1839(2)$ & $\dots$ & $\dots$ & $\dots$ & $\dots$\\[+1ex]\colrule\\[-1ex]
  $N=34$ & $F_\textnormal{xc}/N+\Delta F_\textnormal{xc}$&$ -0.1868(3) $ & $-0.055238(13)$ & $-0.031322(3)$ & $-0.0141743(6)$ & $-0.00756863(14)$\\ 
 & $\Delta F_\textnormal{xc}$& $0.03144 $ & $0.002179$ & $0.00066$ & $0.000134$ & $0.00004$\\[+1ex]
 & $F_\textnormal{F}/N$& $-2.19048(8)$ & $-0.136328(5)$ & $-0.0517098(10)$ & $-0.0174647(2)$ & $-0.00839773(7)$\\
 & $F_{\textnormal{B},0}$ & $-4.10848$ & $-0.16434$ & $-0.0410848$ & $-0.0065736$ & $-0.00164339$\\[+1ex]
 & $S$& $0.2177(2)$ & $0.7467(8)$ & $0.9271(4)$ & $0.99421(5)$ & $0.99956(2)$\\
 & $S_0$& $0.0494(3)$ & $\dots$ & $\dots$ & $\dots$ & $\dots$\\[+1ex]\colrule\\[-1ex]
  $N=50$ & $F_\textnormal{xc}/N+\Delta F_\textnormal{xc}$&$ -0.1861(4) $ & $-0.05523(2)$ & $-0.031323(6)$ & $-0.0141737(8)$ & $-0.0075682(2)$\\ 
 & $\Delta F_\textnormal{xc}$& $0.023842$ & $0.001576$ & $0.000471$ & $0.000094$ & $0.000027$\\[+1ex]
 & $F_\textnormal{F}/N$& $-2.2158(2)$ & $-0.137043(4)$ & $-0.051852(2)$ & $-0.0174770(3)$ & $-0.00839752(9)$\\
 & $F_{\textnormal{B},0}$ & $-4.17724$ & $-0.16709$ & $-0.0417724$ & $-0.0066836$ & $-0.00167089$\\[+1ex]
 & $S$& $0.1035(3)$ & $0.6484(4)$ & $0.8944(3)$ & $0.99128(7)$ & $0.99936(3)$\\
 & $S_0$& $0.0112(2)$ & $\dots$ & $\dots$ & $\dots$ & $\dots$\\[+1ex]\colrule\\[-1ex]
  $N=66$ & $F_\textnormal{xc}/N+\Delta F_\textnormal{xc}$&$ -0.1871(5) $ & $-0.05527(2)$ & $-0.031331(5)$ & $-0.0141746(8)$ & $-0.0075684(2)$\\ 
 & $\Delta F_\textnormal{xc}$& $0.01947 $ & $0.001245$ & $0.000369$ & $0.000073$ & $0.000021$\\[+1ex]
 & $F_\textnormal{F}/N$& $-2.2301(3)$ & $-0.137451(4)$ & $-0.051935(2)$ & $-0.0174852(3)$ & $-0.00839880(9)$\\
 & $F_{\textnormal{B},0}$ & $-4.21564$ & $-0.16863$ & $0.0421564$ & $-0.0067450$ & $-0.00168625$\\[+1ex]
 & $S$& $0.0504(4) $ & $0.5628(6)$ & $0.8623(4)$ & $0.98849(11)$ & $0.99912(4)$\\
 & $S_0$& $0.00237(4)$ & $\dots$ & $\dots$ & $\dots$ & $\dots$\\[+1ex]\colrule\\[-1ex]
 $N=\infty$ & $F_\textnormal{xc}/N$&$-0.1869(3)$ & $-0.055268(9)$ & $-0.031327(3)$ & $-0.014172(3)$ & $-0.007568(3)$\\
 & $F_\textnormal{xc}/N$ (GDSMFB~\cite{groth_prl}) & $-0.1868(6)$ & $-0.0552(2)$ & $-0.031332(10)$ & $-0.01410(5)$ & $-0.00744(3)$\\[1ex]
& $F_\textnormal{F}/N$& $-2.2840(3)$ & $-0.139153(9)$ & $-0.052299(3)$ & $-0.017528(3)$ & $-0.008407(3)$
\end{tabular}
\end{ruledtabular}
\end{table*}

\begin{figure}\centering
\includegraphics[width=0.45\textwidth]{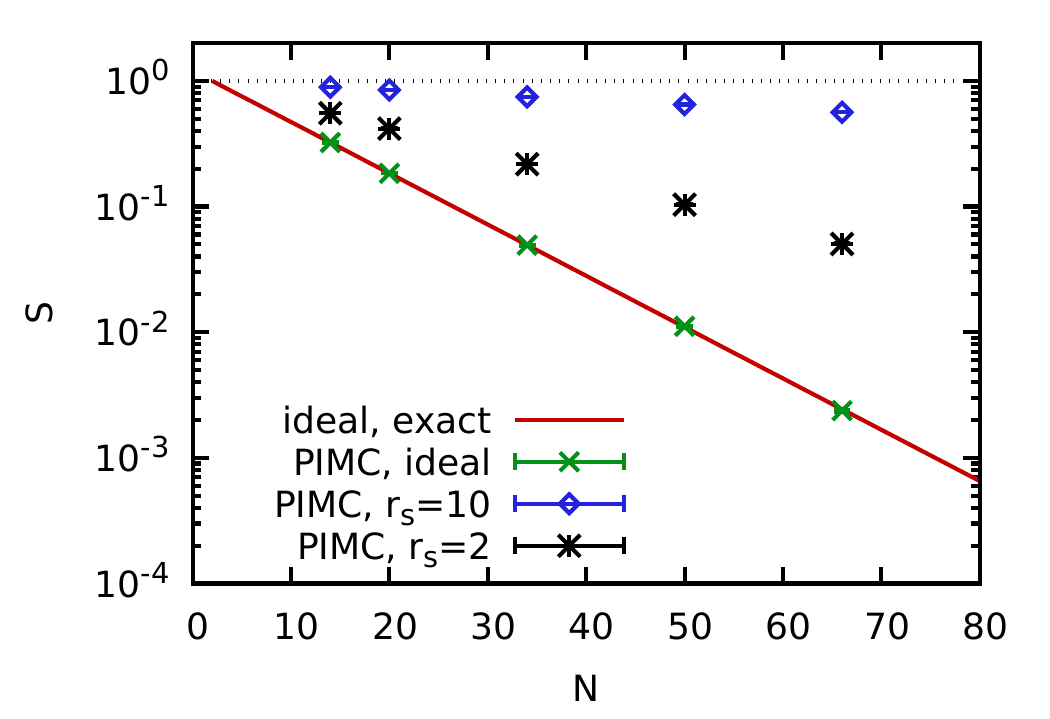}
\caption{\label{fig:sign} Average sign as a function of system size for $\Theta=2$. Solid red: exact results for the ideal Fermi gas computed from Eq.~(\ref{eq:Z_ideal_Bose_Fermi}) via $S_0=Z_{\textnormal{F},0}/Z_{\textnormal{B},0}$; the symbols show PIMC results for the ideal Fermi gas (green crosses), as well as for the UEG at $r_s=2$ (black stars) and $r_s=10$ (blue diamonds).
}
\end{figure} 

\begin{figure*}\centering
\includegraphics[width=0.33\textwidth]{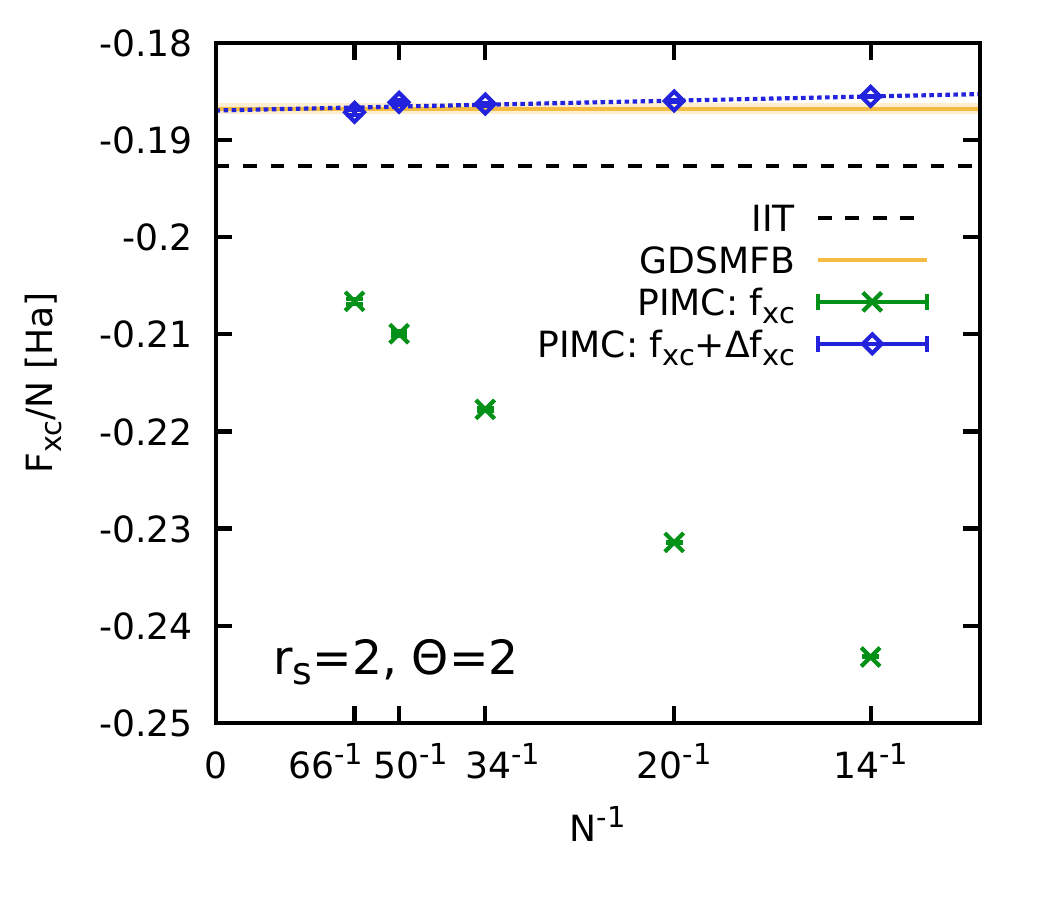}\includegraphics[width=0.33\textwidth]{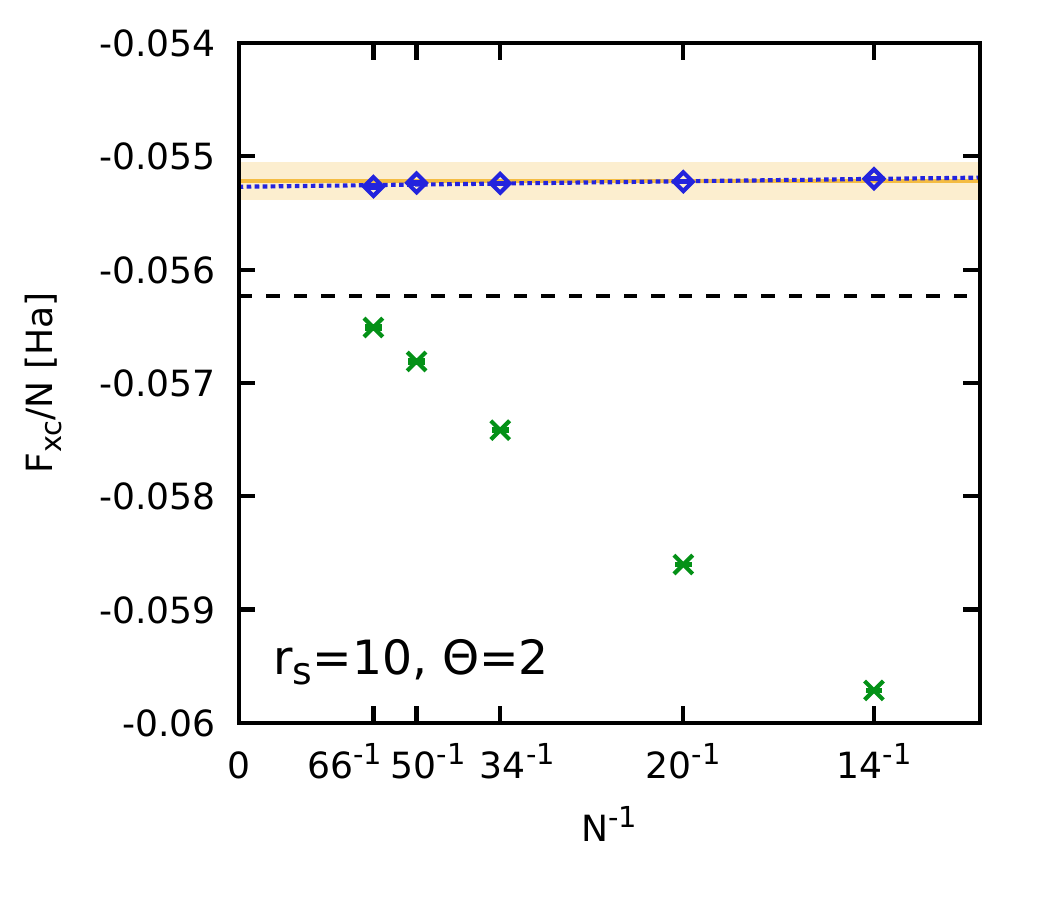}\includegraphics[width=0.33\textwidth]{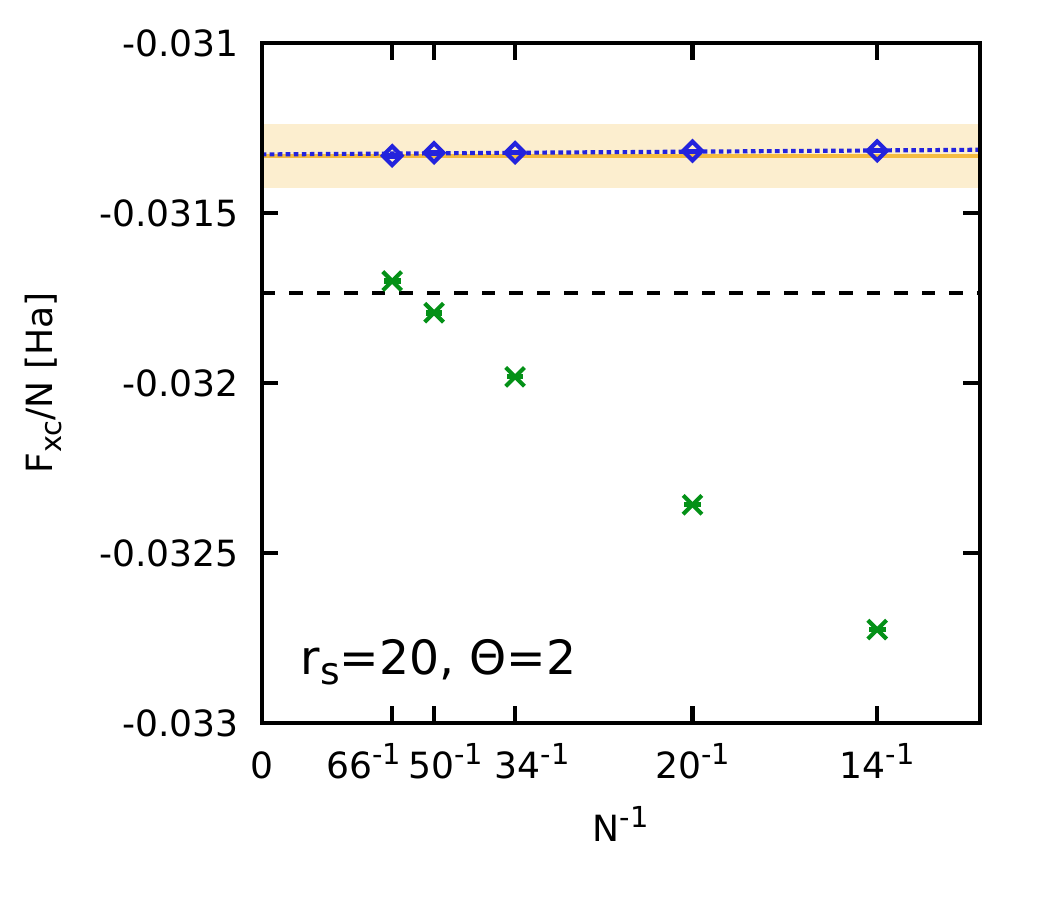}\\\includegraphics[width=0.33\textwidth]{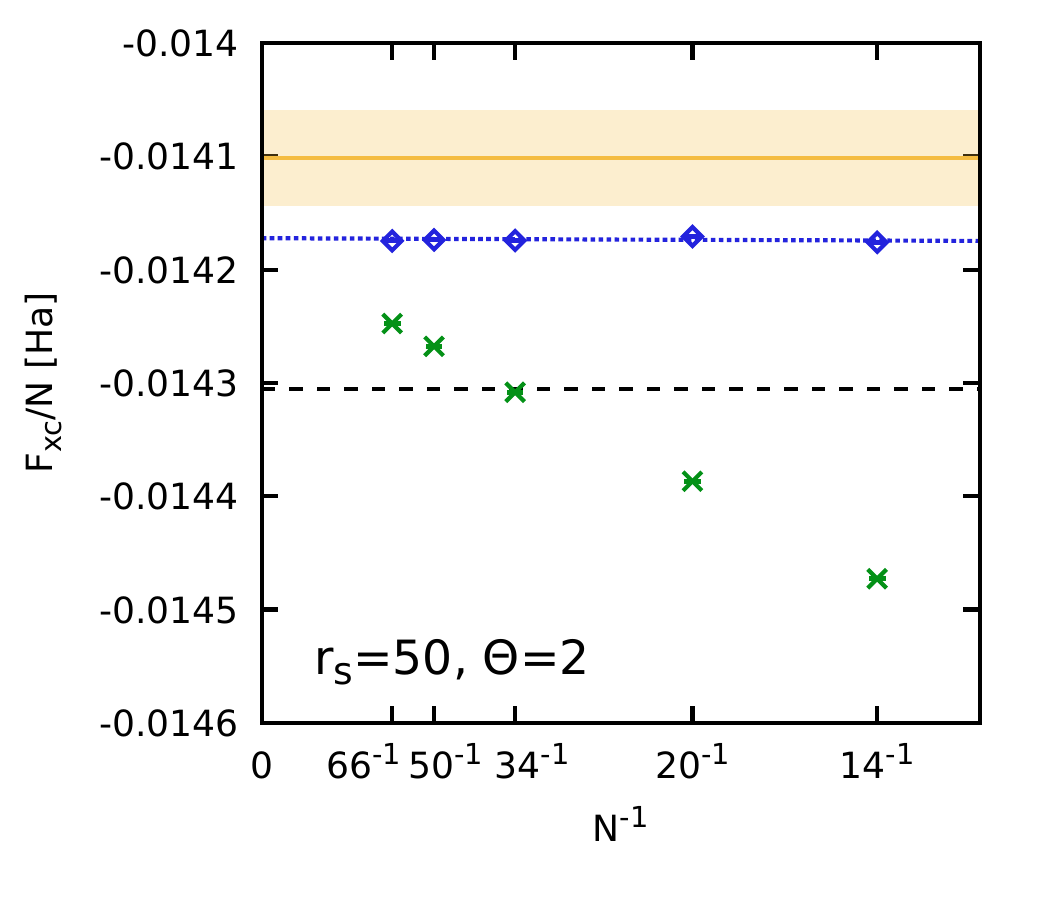}\includegraphics[width=0.33\textwidth]{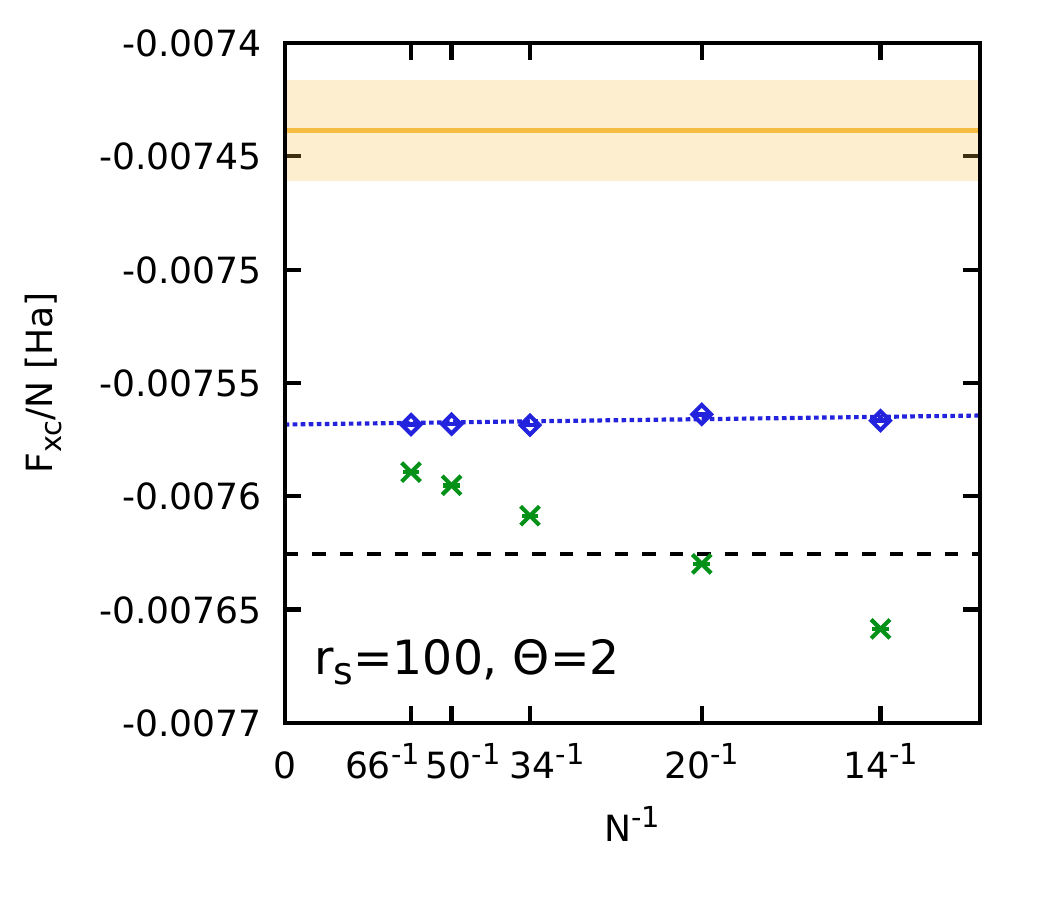}
\caption{\label{fig:xc} 
The exchange--correlation free energy of the UEG at $\Theta=2$ and different values of the density parameter $r_s$. Green crosses: raw PIMC results [Eq.~(\ref{eq:Fxc})]; blue diamonds: finite-size corrected PIMC results (with $\Delta F_\textnormal{xc}$ computed using the \texttt{UEGPY} code~\cite{uegpy}); dotted blue line: empirical linear extrapolation of the small residual system-size dependence; horizontal dashed black line: parametrization by Ichimaru, Iyetomi and Tanaka (IIT)~\cite{IIT}; horizontal yellow line (shaded yellow area): parametrization by Groth \emph{et al.}~\cite{groth_prl} (GDSMFB) and corresponding uncertainty interval of $\pm0.3\%$. Note that the nominal validity range of GDSMFB is given by $r_s\leq 20$. The PIMC results for $r_s=2$ are partially taken from Ref.~\cite{dornheim2024directfreeenergycalculation}.
}
\end{figure*} 

\begin{figure*}\centering
\includegraphics[width=0.33\textwidth]{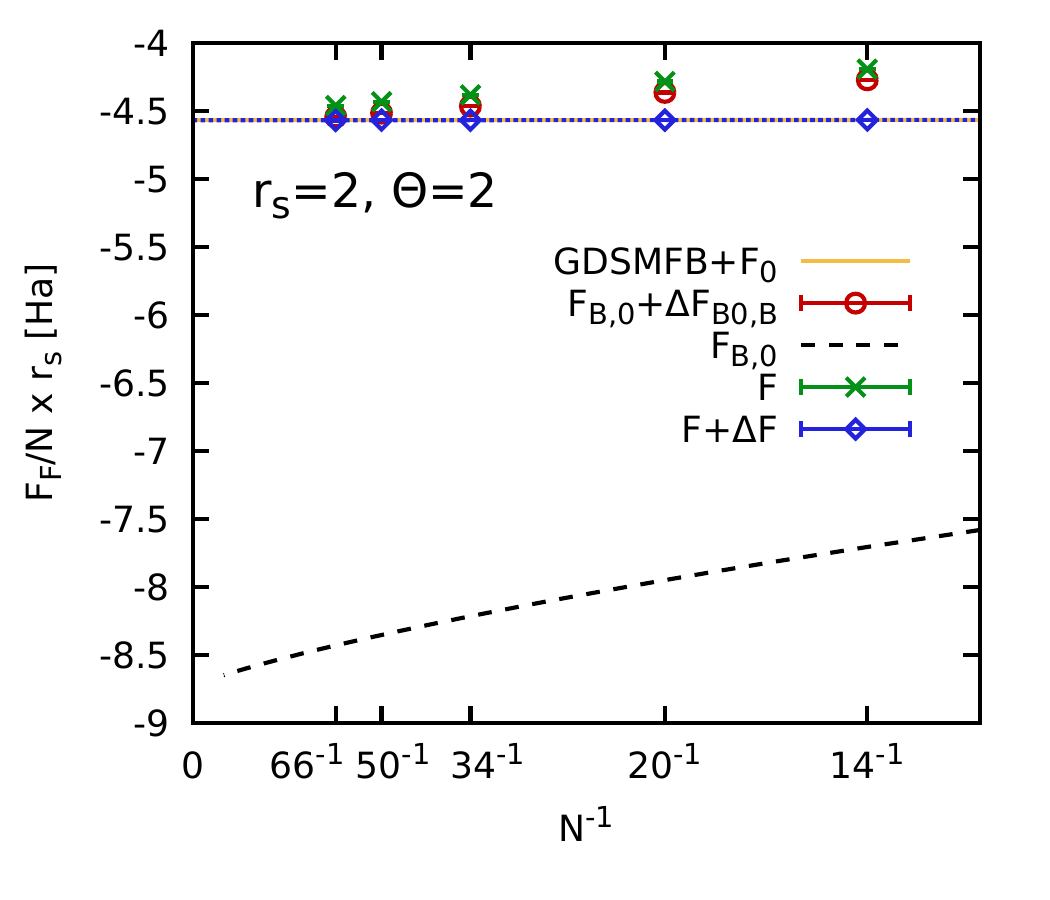}\includegraphics[width=0.33\textwidth]{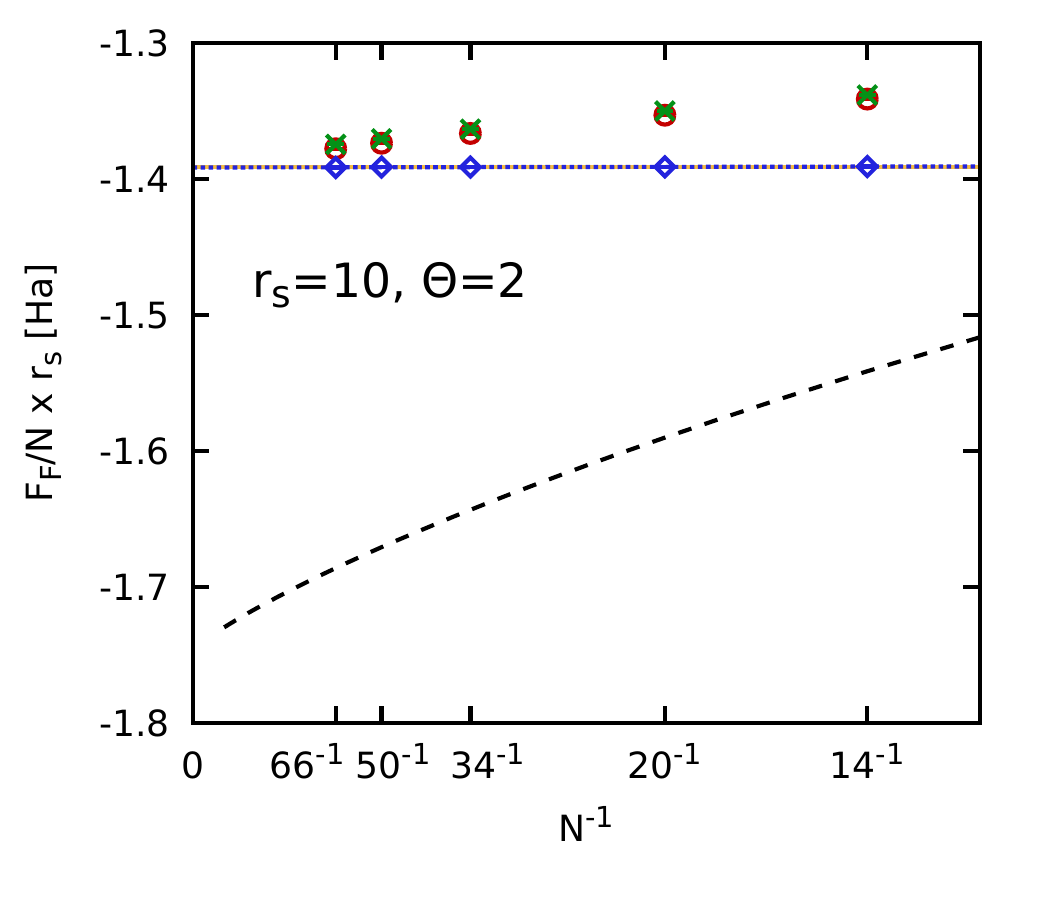}\includegraphics[width=0.33\textwidth]{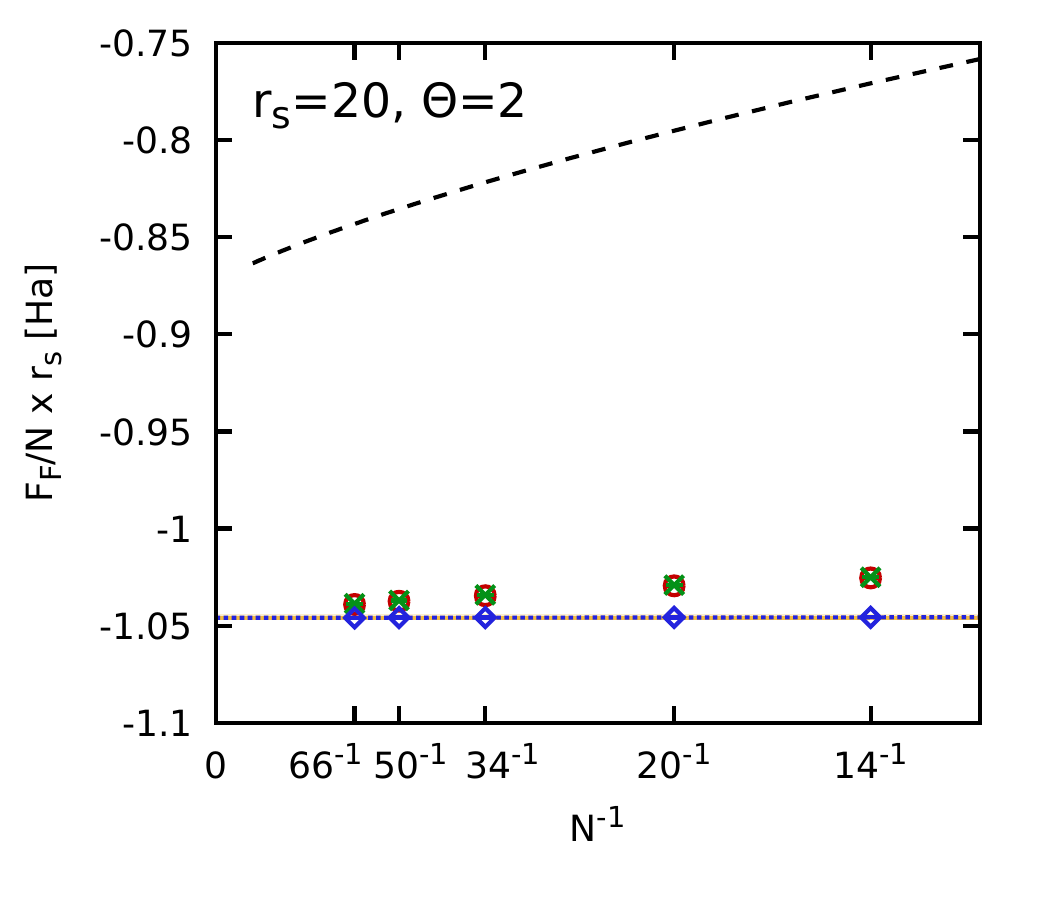}
\includegraphics[width=0.33\textwidth]{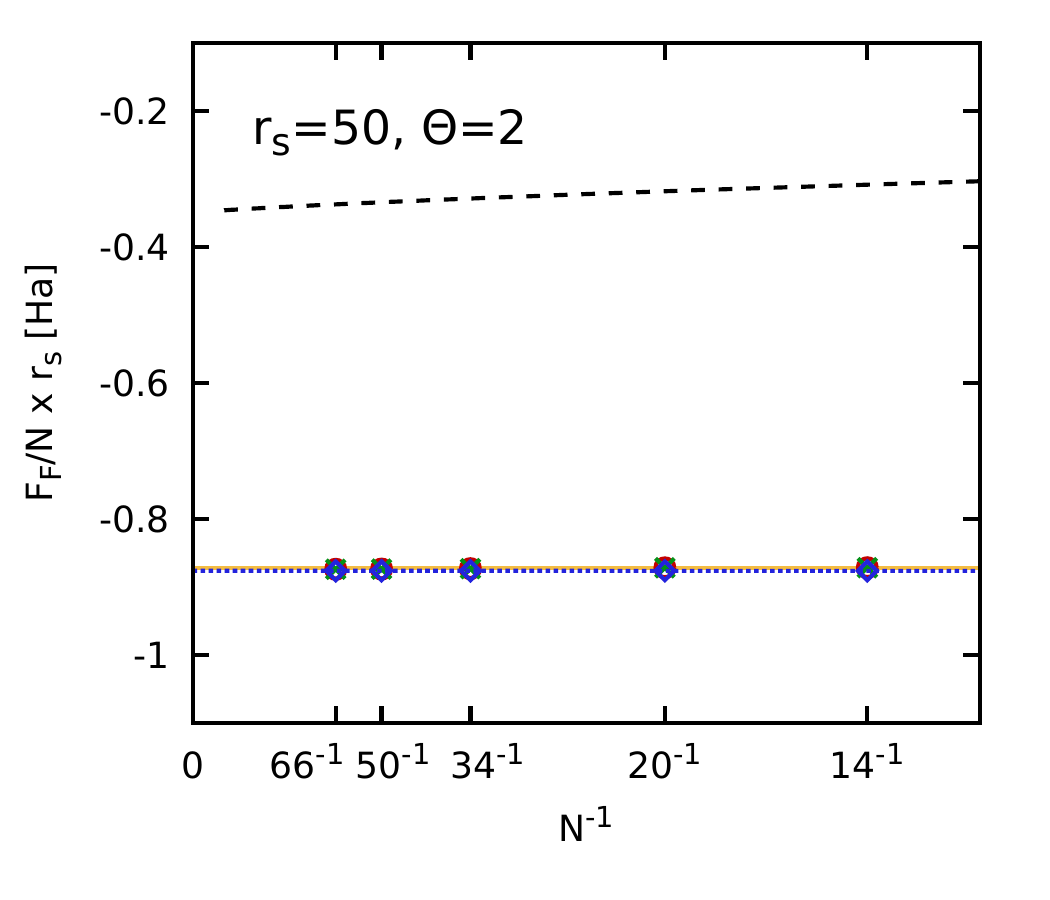}\includegraphics[width=0.33\textwidth]{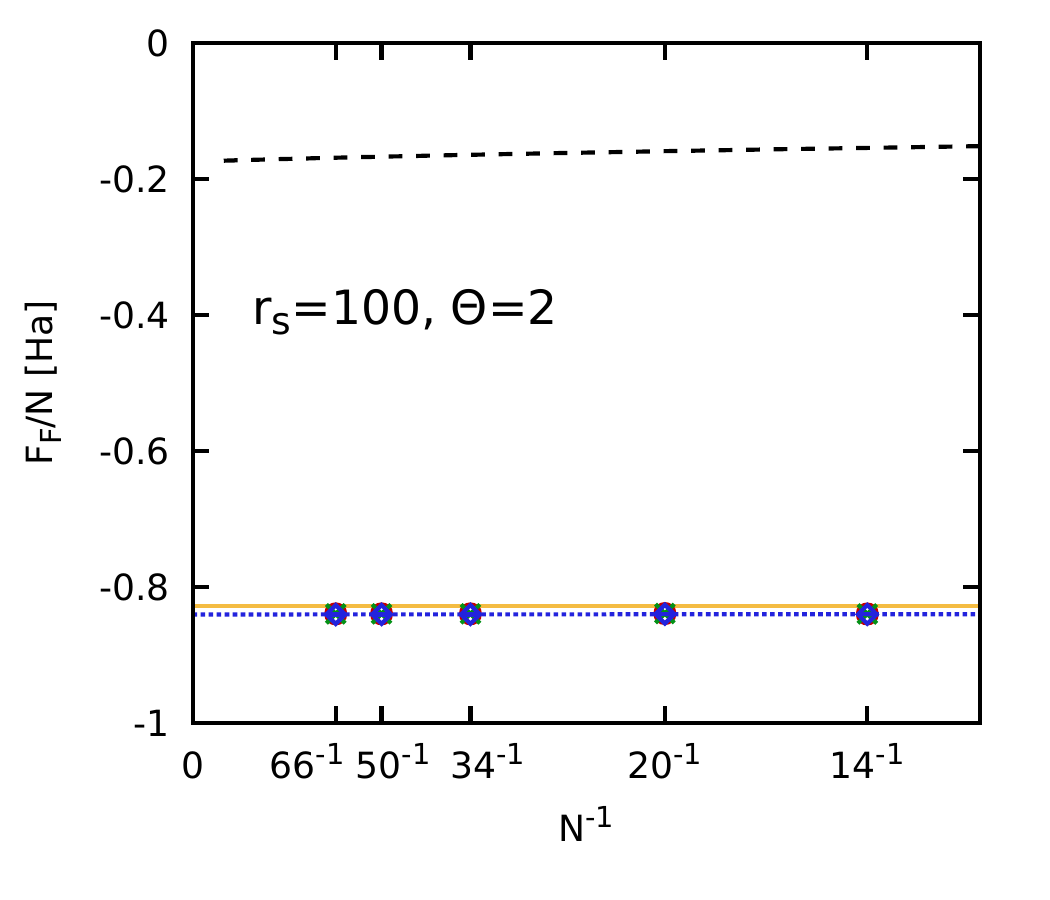}
\caption{\label{fig:total} 
The total free energy of the UEG at $\Theta=2$ and different values of the density parameter $r_s$. Green crosses: raw PIMC results [Eq.~(\ref{eq:final})]; red circles: PIMC results without the quantum statistical correction $\Delta F_{F,B}$ [cf.~Eq.~(\ref{eq:decomposition})]; blue diamonds: finite-size corrected PIMC results [Eq.~(\ref{eq:FSC_total})]; dotted blue line: linear fit to the blue diamonds; dashed black line: free energy of the noninteracting Bose reference system $F_{\textnormal{B},0}$;
horizontal yellow line (shaded yellow area): parametrization by Groth \emph{et al.}~\cite{groth_prl} (GDSMFB) (corresponding nominal uncertainty of $\pm0.3\%$ of $f_\textnormal{xc}$).
}
\end{figure*} 

Let us next investigate the exchange--correlation contribution to the total free energy $F_\textnormal{xc}$, which is of key importance for a number of practical applications, most importantly as the basis for thermal DFT simulations~\cite{wdm_book}. For the UEG, there is no Hartree contribution, thus, we have $F_\textnormal{xc}\equiv F_\textnormal{F}-F_{\textnormal{F},0}$, where $F_{\textnormal{F},0}$ denotes the free energy of the ideal Fermi gas. In the thermodynamic limit $N,\Omega\to\infty$, the estimation of $F_{\textnormal{F},0}$ is trivial. For a finite particle number $N$, $F_{\textnormal{F},0}$ can be estimated from recursion relations of the canonical partition function~\cite{PhysRevResearch.2.043206,DuBois,Zhou_2018}. Alternatively, we can again use the free energy of the non-interacting Bose gas as a basis and use the corresponding average sign $S_0=Z_{\textnormal{F},0}/Z_{\textnormal{B},0}$ to compute the respective quantum statistical correction,
\begin{eqnarray}\label{eq:Fxc}
    F_\textnormal{xc} = F_\textnormal{F} - \underbrace{F_{\textnormal{B},0} -  \frac{1}{\beta}\textnormal{log}\left( S_0
    \right)}_{F_{\textnormal{F},0}}\ .
\end{eqnarray}
In Fig.~\ref{fig:sign}, we show $S_0$ computed from the compact recursion relation [Eq.~(\ref{eq:Z_ideal_Bose_Fermi})] as the solid red curve; it is in excellent agreement with the green crosses that have been obtained from corresponding PIMC simulations of the ideal Fermi gas. Evidently, the sign exhibits an exponential decay with $N$, which becomes increasingly steep for lower temperatures. Even for $\Theta=2$ and $N=66$, $S_0=0.00237(4)$ is hard to resolve with PIMC. This makes the direct evaluation of the recursion relation Eq.~(\ref{eq:Z_ideal_Bose_Fermi}) the preferred option, although it, too, becomes numerically unstable for large $N$ and/or low $\Theta$.
The black crosses and blue diamonds in Fig.~\ref{fig:sign} show PIMC results for the average sign of the interacting UEG at $r_s=2$ and $r_s=10$, respectively. As it is expected~\cite{dornheim_sign_problem}, these data sets, too, exhibit exponential decays with $N$, but much less steep compared to the noninteracting case. This nicely illustrates the advantages of carrying out the $\eta$-connection in the bosonic sector, avoiding the substantially more severe sign problem of the ideal Fermi gas, cf.~Fig.~\ref{fig:scheme} above.

In Fig.~\ref{fig:xc}, we show our new PIMC results for the exchange--correlation free energy per particle $f_\textnormal{xc}=F_\textnormal{xc}/N$ over a broad range of densities starting at WDM conditions ($r_s=2$, top left) and going all the way to the strongly coupled electron liquid regime ($r_s=100$, bottom right). The green crosses correspond to our raw PIMC results [Eqs.~(\ref{eq:final},\ref{eq:Fxc})], and we find a substantial dependence on the system size, which, in turn, strongly depends on the density. For $r_s=2$, finite-size effects are most pronounced and attain $\sim25\%$ for $N=14$ electrons. In contrast, our PIMC results only vary by about $\sim1\%$ for $r_s=100$. This is the expected trend, which is in good agreement with the existing literature~\cite{dornheim_prl,review,Dornheim_JCP_2021,dornheim_electron_liquid}. The blue diamonds have been obtained by adding a finite-size correction $\Delta F_\textnormal{xc}$ using the approach introduced in Refs.~\cite{Drummond_PRB_2008,Brown_PRL_2013,dornheim_prl}, see also the Supplemental Material of Ref.~\cite{dornheim2024directfreeenergycalculation} for its particular application to $f_\textnormal{xc}$, that has been implemented into the \texttt{UEGPY} code by F.D.~Malone~\cite{uegpy}.
Evidently, adding the finite-size correction removes the bulk of the system-size dependence, and small residual finite-size effects are removed with a subsequent empirical linear extrapolation (dotted blue).

\begin{figure*}\centering
\includegraphics[width=0.33\textwidth]{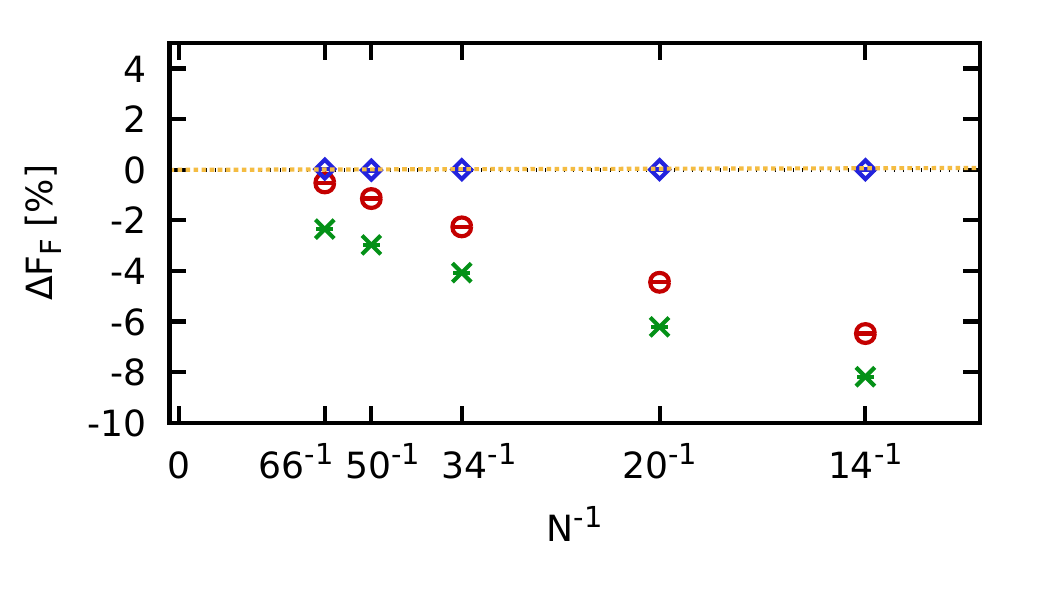}\hspace*{0.00\textwidth}\includegraphics[width=0.33\textwidth]{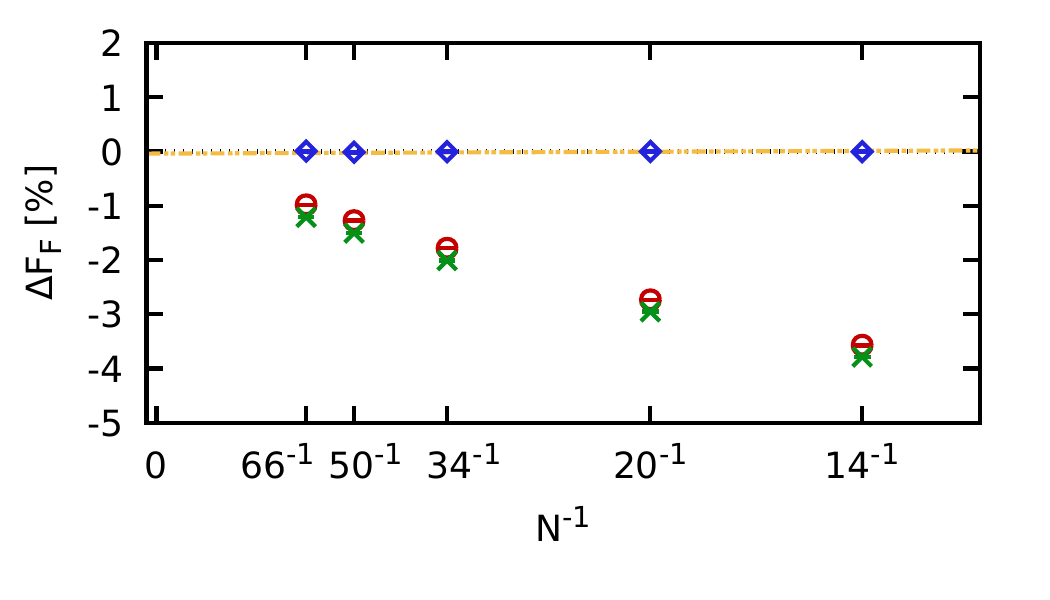}\includegraphics[width=0.33\textwidth]{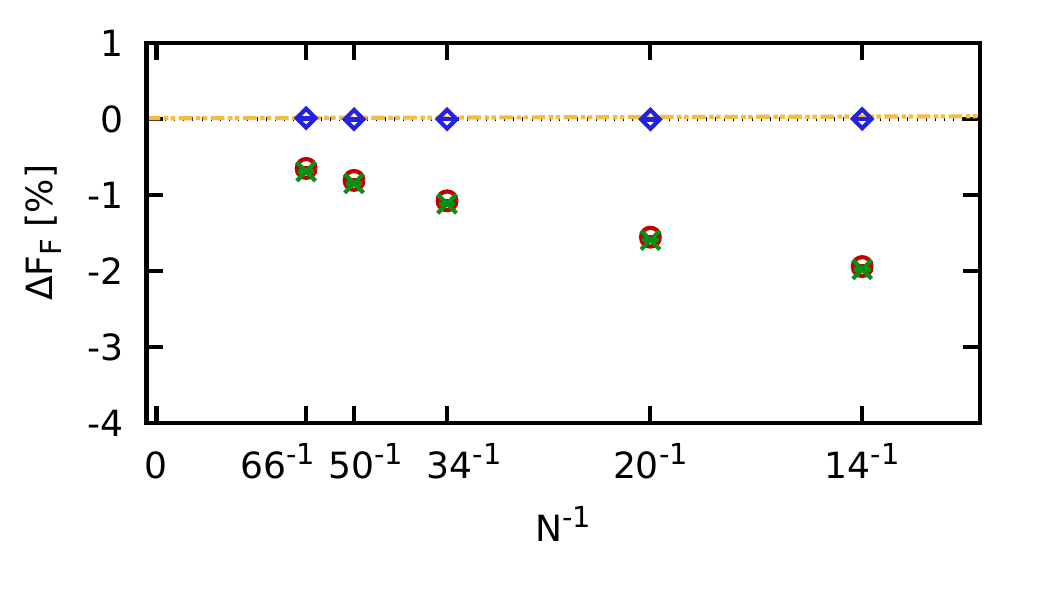}\\
\includegraphics[width=0.33\textwidth]{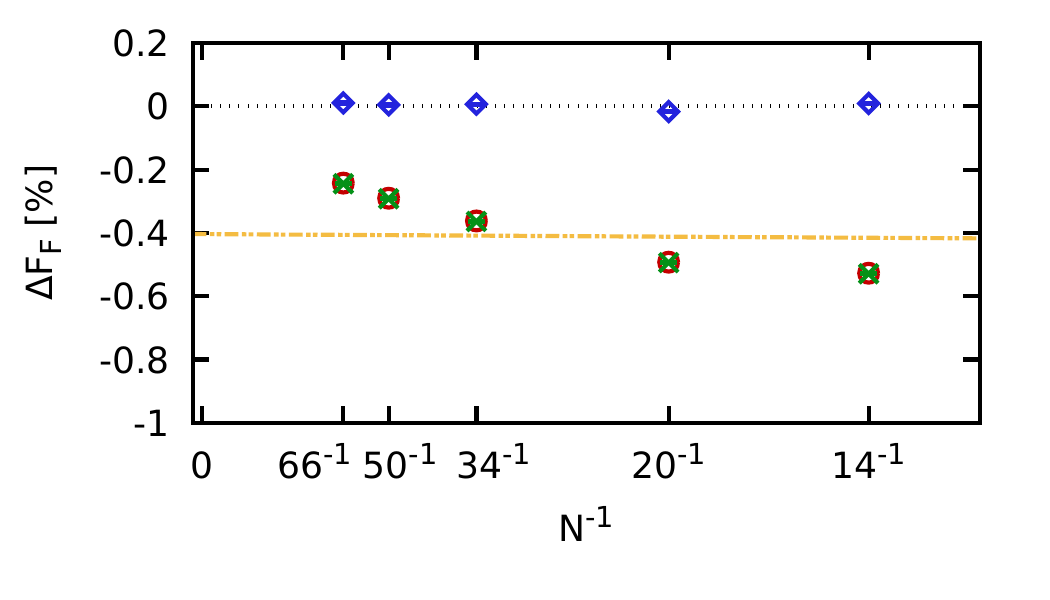}\includegraphics[width=0.33\textwidth]{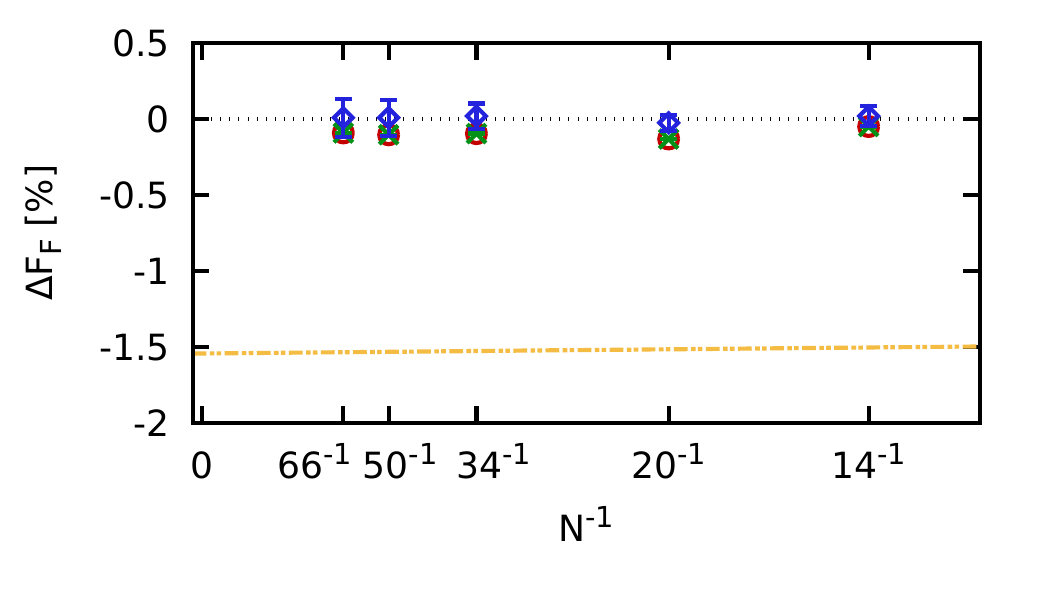}
\caption{\label{fig:delta_total} The relative deviation to the linear fits of finite-size corrected PIMC results for the total free energy $F_\textnormal{F}/N$ of the UEG shown in Fig.~\ref{fig:total}:
Green crosses: raw PIMC results [Eq.~(\ref{eq:final})]; red circles: PIMC results without the quantum statistical correction $\Delta F_{F,B}$ [cf.~Eq.~(\ref{eq:decomposition})]; 
blue diamonds: finite-size corrected PIMC results [Eq.~(\ref{eq:FSC_total})];
dotted yellow line: parametrization by Groth \emph{et al.}~\cite{groth_prl} (GDSMFB).
}
\end{figure*} 

For $2\leq r_s \leq 20$ (top row of Fig.~\ref{fig:xc}), we find excellent agreement with the parameterization of $f_\textnormal{xc}(r_s,\Theta)$ by Groth \emph{et al.}~\cite{groth_prl} (GDSMFB, yellow line), which has been obtained from a thermodynamic integration over PIMC results for the interaction energy; the shaded yellow area indicates the nominal uncertainty of $\pm0.3\%$ of their representation. In contrast, our results have been obtained solely from a PIMC simulation for the respective density--temperature combination. The conditions shown in the bottom row of Fig.~\ref{fig:xc} have been obtained for $r_s=50$ and $r_s=100$, which are significantly outside of the range of applicability of the GDSMFB parametrization. Still, we find that the latter reproduces the correct qualitative behavior and that it remains accurate to $\sim2\%$ even for $r_s=100$. The results are also compared to the parameterization of $f_\textnormal{xc}(r_s,\Theta)$ by Ichimaru, Iyetomi and Tanaka (IIT)~\cite{IIT,Tanaka_CPP_2017} (IIT, dashed black line), which has been obtained from a thermodynamic integration of an STLS-based interaction energy expression modified to ensure that the exact ground state limit $\Theta\to0$ (as obtained from QMC simulations~\cite{Ceperley_Alder_PRL_1980}) and the exact classical limit $\Theta\to\infty$ (as obtained from MC simulations~\cite{PhysRevA.21.2087}) are followed. The IIT parametrization becomes more accurate than the GDSMFB parametrization only for $r_s=100$. This is a direct consequence of the incorporation of exact strong coupling results at the classical limit.

\subsection{Total free energy\label{sec:total}}

In Fig.~\ref{fig:total}, we show the total free energy $F_\textnormal{F}$ of the UEG. In this case, the evaluation of Eq.~(\ref{eq:final}) by itself does not require us to resolve the average sign of the degenerate ideal Fermi gas, but only the sign of the interacting UEG. We find $S\sim0.05$ even for the most difficult case of $N=66$ at $r_s=2$ and $\Theta=2$, which means that the simulations are involved, but certainly feasible. In contrast, we have $S\approx1$ for $r_s=100$ as the formation of permutation cycles is effectively suppressed by the strong Coulomb repulsion in the electron liquid regime~\cite{Dornheim_permutation_cycles}.

The dashed blue lines show the free energy of a non-interacting Bose gas at the same conditions. Adding the bosonic interacting correction $\Delta F_{\textnormal{B},0}$ [cf.~Eq.~(\ref{eq:decomposition})] gives the red circles, and subsequently adding the quantum statistical correction $\Delta F_{\textnormal{F},\textnormal{B}}$ results in the green crosses that constitute our final result for the free energy of the $N$-electron UEG.
Clearly, Coulomb coupling effects play a dominant role and decisively impact the final result. In contrast, the quantum statistical correction computed from the average sign is less important here and, even at $r_s=2$, only constitutes $\sim2\%$ of the total free energy. To obtain a suitable finite-size correction for the PIMC data for $F_\textnormal{F}/N$, it is convenient to consider the previously discussed decomposition into an XC part and a residual non-interacting contribution,
\begin{eqnarray}\label{eq:FSC_total}
    F_\textnormal{F} = F_{\textnormal{F},0} + F_\textnormal{xc} \Rightarrow \Delta F_\textnormal{F} = \Delta F_\textnormal{xc} + \Delta F_{\textnormal{F},0}\ .
\end{eqnarray}
The computation of $\Delta F_{\textnormal{F},0}$ is straightforward,
\begin{eqnarray}\label{eq:Delta_F_0}
    \Delta F_{\textnormal{F},0}(N) = f_{\textnormal{F},0}(\infty) - F_{\textnormal{F},0}(N)/N\ ,
\end{eqnarray}
as both contributions can be directly evaluated, see also Eq.~(\ref{eq:Fxc}) above. The thus finite-size corrected PIMC results are shown as the blue diamonds in Fig.~\ref{fig:total}. Evidently, the dependence on the system size has been drastically reduced, and no residual size dependence can be resolved on the depicted energy scale. The dotted blue lines show the corresponding empirical fits to the corrected data set, but the linear coefficient is hardly significant for all considered parameters. This can be discerned particularly well in Fig.~\ref{fig:delta_total}, where we show the relative difference between the different data sets and the fit. First, no residual finite-size effects can be resolved; second, finite-size effects are maximal for $r_s=2$ and attain $8\%$ for $N=14$, whereas they are negligible in practice at $r_s=100$; this is expected to change for even lower densities when the UEG starts to crystallize and commensurability effects start to become important~\cite{Clark_PRL_2009}; third, the effects of quantum statistics barely attain $0.1\%$ even for $r_s=10$ at the investigated temperature; investigating the effects of quantum statistics in the low-temperature and low-density UEG constitutes an interesting topic for dedicated future research.

All PIMC results are listed in Table~\ref{tab:UEG}.

\section{Summary and Outlook\label{sec:outlook}}

In this work, we have presented an in-depth investigation of the recently introduced~\cite{dornheim2024directfreeenergycalculation} $\eta$-ensemble approach to the free energy of quantum many-body systems applied to the UEG.
In the first part of our study, we have analyzed the efficiency and ergodicity of the approach with respect to the number of $\eta$-steps $N_\eta$ and with respect to the free weighting parameter $c_\eta$. As it can be expected, connecting the ideal reference system ($\eta=0$) with the interacting physical system of interest ($\eta=1$) becomes more difficult for strong coupling, i.e., towards low densities and low temperatures. The free parameter $c_\eta$ is important to balance constant energy shifts between the two considered $\eta$-values (e.g., the constant energy contribution from the uniform positive background in the case of the UEG), but it becomes insufficient when the overlap between the respective configuration spaces becomes too small. In that case, one has to increase the number of $\eta$-steps, which makes it always possible to restore the sampling efficiency of both sectors, and transitions in between. In practice, we find that ergodicity is most problematic in the limit of small $\eta$ and, in particular, for transitions between $\eta_2=0$ and a finite value for $\eta_1$. This is a consequence of the absent energetic punishment for two very close particles in the ideal limit, whereas such configurations are punished exponentially with decreasing particle distance for any finite value of $\eta$. Consequently, we recommend a finer $\eta$-grid for small $\eta$, whereas larger transitions are empirically unproblematic for $\eta\gtrsim0.5$. A second, a priori less obvious empirical observation concerns the decreasing sampling efficiency for transitions between two $\eta$-sectors upon increasing the system size. Here, too, path configurations with near particles linearly increase in likelihood with increasing $N$ for $\eta=0$, but exponentially decrease the configuration weight for any finite value of $\eta$; again, the problem can be easily solved by increasing the number of $\eta$-steps $N_\eta$, in particular in the limit of small $\eta$.

In the second part of our work, we have carried out extensive new PIMC simulations to estimate the exchange--correlation free energy $F_\textnormal{xc}$ and total free energy $F$ of the UEG over a broad range of densities extending from the WDM regime ($r_s=2$) to the strongly coupled regime ($r_s=100$). Clearly, the approach works well for all considered parameters. For $r_s\leq20$, we have found excellent agreement with the existing parametrization of $F_\textnormal{xc}/N$ by Groth \emph{et al.}~\cite{groth_prl}, which is based on the thermodynamic integration of extensive quantum Monte Carlo results for the interaction energy $V$ for a large set of densities and temperatures; in contrast, our results have been obtained exclusively based on PIMC calculation for a single density--temperature combination in each case. For $r_s=50$ and $r_s=100$, no reliable PIMC results had been available. Still, we find that the GDSMFB parametrization works reasonably well (with a systematic error in $F_\textnormal{xc}/N$ of $\sim1.5\%$ for $r_s=100$), well outside of its intended range of application.

We have further investigated the total free energy of the UEG, which allows us to analyze its natural decomposition into a free bosonic contribution, a bosonic interaction contribution, and a quantum statistical contribution. Interestingly, the latter is comparably small even at $r_s=2$, where it attains $\sim2\%$. In addition, the bosonic interaction contribution becomes more important for lower densities, as it is expected. Finally, we have found that PIMC results for the free energy of the UEG can be reliably finite-size corrected by combining the corresponding corrections of the XC-free energy and the ideal Fermi gas.

Future works include the further optimization of the $\eta$-ensemble approach by implementing an improved handling of Coulomb interactions in the Monte Carlo sampling~\cite{PhysRevX.13.031006} or path contractions for large systems~\cite{PhysRevE.93.043305}. Moreover, its application to electron--nuclei systems such as warm dense hydrogen~\cite{Dornheim_MRE_2024} and beryllium~\cite{Dornheim_Science_2024} will allow for rigorous benchmarks of existing equation-of-state tables~\cite{Militzer_PRE_2021}, which constitute key input for astrophysical models and IFE predictions alike~\cite{drake2018high}. Furthermore, the $\eta$-ensemble approach can easily be applied to inhomogeneous systems~\cite{dornheim2024directfreeenergycalculation} such as the electronic problem in the external potential of a fixed ion configuration, which will be of high value to benchmark DFT calculations with different XC-functionals. Finally, the approach can also be employed for UEG phase diagram calculations at finite temperatures. Calculations of the crystallization line are particularly challenging for any system due to the tiny energy differences between the competing phases in the vicinity of the phase transition and would certainly benefit from the elimination of the need for thermodynamic integration. It is emphasized that there are no available predictions for the crystallization line of the finite temperature UEG and that there is no consensus even for the crystallization density of the extensively studied ground state limit~\cite{Drummond_PRB_Wigner_2004,Holzmann_PRL_2020,Azadi_PRB_2022}.

\begin{acknowledgements}

\noindent This work was partially supported by the Center for Advanced Systems Understanding (CASUS), financed by Germany’s Federal Ministry of Education and Research and the Saxon state government out of the State budget approved by the Saxon State Parliament. This work has received funding from the European Research Council (ERC) under the European Union’s Horizon 2022 research and innovation programme (Grant agreement No. 101076233, "PREXTREME"). 
Views and opinions expressed are however those of the authors only and do not necessarily reflect those of the European Union or the European Research Council Executive Agency. Neither the European Union nor the granting authority can be held responsible for them. Computations were performed on a Bull Cluster at the Center for Information Services and High-Performance Computing (ZIH) at Technische Universit\"at Dresden and at the Norddeutscher Verbund f\"ur Hoch- und H\"ochstleistungsrechnen (HLRN) under grant mvp00024.
\end{acknowledgements}

\bibliography{bibliography}
\end{document}